%% file: ett2.6.tex
\documentclass[10pt]{article}

\usepackage[margin=1.5in]{geometry}

\input{amslambda.tex}

\usepackage{url}
\usepackage{relsize}
\usepackage{graphicx}
\usepackage[dvipsnames]{xcolor}
\usepackage{rotating}
\usepackage{amsthm}

\renewcommand{\bsim}{{\raisebox{-1mm}{\scalebox{1.5}{$\sim$}}}}

\newenvironment{lemma}{{\sc Lemma.}}{}
\newenvironment{definition}{{\sc Definition.}}{}
\newenvironment{theorem}{{\sc Theorem.}}{}

\title{Internalization of extensional equality}
\author{Andrew Polonsky}

\usepackage{ebharvard}

\usepackage{titlesec}
\titlelabel{\thetitle.\quad}

\renewcommand{\lext}{\leeq}
\renewcommand{\optarg}[1]{\greyout{#1}}

\newcommand{\iso}{\stackrel{\eeq}{\to}}
  \newcommand{\phabra}{\phantom{\{}}

\newcommand{\objof}[1]{\mathsf{Ob}({#1})}
\newcommand{\homsym}{\mathsf{Hom}}
\begin{document}
\maketitle

\begin{abstract}
We give a type system in which the universe of types is closed by
reflection into it of the logical relation defined
externally by induction on the
structure of types.
This contribution is placed in the context of the search for a
natural, syntactic construction of the extensional equality type
(\cite{tait}, \cite{thorsten1999}, \cite{coquand}, \cite{harper}, \cite{perML}).
The system is presented as an extension of $\lstar$, the terminal pure
type system in which the universe of all types is a type.
The universe inconsistency is then removed by the usual method of
stratification into levels.
We give a set-theoretic model for the stratified
system.  We conjecture that Strong Normalization holds as well.
\end{abstract}

\section{Background: the problem of extensionality}

In recent years, the problem of extensionality in type theory has
received increasing attention.  In part, this is due to
type theory emerging as the language of choice for
computer formalisation of mathematics.
(\cite{Gonthier},  \cite{thehomotopybook}.)

The fundamental notion of this language, that of a \emph{type},
is a notion of collection which bases membership on the
syntactic form of the objects.  Accordingly, the notion of
equality between objects of a given type is likewise based
on their syntactic form:
two expressions are judged as denoting equal objects
if one can be transformed into another by a finite
sequence of syntactic manipulations.

Since a general number-theoretic function can in principle be
implemented in any number of ways, there will be
different expressions defining the same function which cannot
be transformed from one to another using syntactic manipulations only.
For example, the function which maps a vector of numbers
to a rearrangement of it listing the numbers in non-decreasing order,
can be implemented using bubble-sort or quick-sort processes,
and these cannot be transformed into one another by local
simplifications.

Therefore, basing equality on syntactic form alone leads to the
failure of \emph{function extensionality}, the principle stating that
two functions are equal if they are pointwise equal:
\[ \forall x,y\ (x=_A y \RA fx =_B gy)\qquad \then \qquad f=_{A \to B}
g \tag{FE} \]
At the same time, this principle is deeply embedded into the
language and culture of mathematics.  After set theory
became the standard language of 
communicating mathematical ideas, the intuitive notion of ``function''
came to be understood through its encoding into
set theory ---  as a set of ordered pairs.
Since sets are extensional almost by definition, so must be functions in
the set-theoretic foundations.

In the usual mathematics, the above ``principle'' is therefore a
matter of linguistics.
\footnote{At the time of
writing these words, the definition of the word ``function'' given by
Google is that it is an ``expression with one or more variables''.
This corresponds to the type-theoretic notion of a function ---  a lambda term ---
and not the set-theoretic notion (a set of pairs having certain
properties).

The middle-school definition of a function as a ``black box''
which \emph{transforms} its input into output is
likewise more faithfully captured by the lambda calculus standpoint.
}

But not so in type theory. Elementary mathematical arguments often become
clumsy when translated to type theory, because the principle (FE) is
not available.

In order to develop set-theoretic mathematics in type theory,
it would be convenient to have a notion of equality that
justified the above principle.  Such a notion may then be called
\emph{extensional equality}.  Unfortunately, the known constructions
of this notion result in violation of key design principles of type theory.

\section{Approaches to extensionality}

The classical approach originally pursued by \cite{itt} is
to extend the definitional (syntactic) equality between expressions,
by allowing expressions to be declared syntactically equal whenever
the statement of their equality is proved in the system's logic.
(That is, mathematical equality is reflected back into the syntax.)

This fixed the problem with (FE), but the cost was too great:
syntactic equality, being now dependent on propositions, became
undecidable, and so did type-checking.  But type theory, following
Curry--Howard paradigm, identifies proving propositions with
inhabiting types.  Thus undecidability of type-checking means it is
not possible to decide whether a purported proof is indeed a proof,
making type theory useless as a foundational system. \footnote{
After all, the point of a formal system is not to abstractly talk
about the set
\[ \setof{A \mid A \text{ is true}} \]
but rather to offer \emph{convenient} tools whereby membership (of some elements)
in this set can be established with finite effort.}

A more recent idea is due to \cite{hlc},
who discovered a single sentence in the language of type theory
which, when assumed as an axiom, makes the intensional identity type
behave like the extensional one.  The axiom is a strong form of \emph{universe
extensionality} (\cite{groupoid}) --- stating that isomorphic types
are equal --- and has many deep consequences for type theory,
including (FE).

However, assuming an axiom in type theory without specifying
its combinatory behavior with respect to other symbols
results in the loss of another crucial property of \emph{canonicity}.
This property guarantees that every type-theoretic construction
is conceptually computable, in the sense that every definition can
be computed by trivial simplification steps to a value.

A computational interpretation for Voevodsky's axiom has been recently
given by Coquand and collaborators (\cite{cubes}).  Their solution is
\emph{semantic}:
all relevant computations are performed in
a constructive model of homotopy types.
The notion of extensional equality is given meaning only implicitly,
via its interpretation in the homotopy model of type theory.

It would appear worthwhile to have a native, type-theoretic construction
of extensional equality which did not assume the univalence axiom or
the homotopy interpretation.  

\section{Extensional equality by induction on type structure}

Yet another approach is to define
extensional equality by induction on type structure.
Here one defines the equality relation
\emph{externally} to the type system, by relating
certain elements of the (free) term model of the system.

This is actually the oldest approach to the problem,
having led \cite{Gandy} to derive the
\emph{logical relations principle}, a basic tool in metatheoretic
studies of type systems.

In 1995, in a paper titled ``Extensional Equality in the Classical
Theory of Types'', William Tait assumed the denotation of the notion of
extensional equality to be the canonical equivalence relation
defined by the logical relations principle.  While this idea is
certainly familiar to many researchers,
it appears to have missed a general announcement.

Let it thus be made explicit. \vspace{0.3cm}

{\centering \framebox{
\begin{minipage}{11cm}
{\sc \ul{Extensionality Thesis}.} \emph{The extensional equality type
  is the canonical equivalence 
  relation defined between elements of the term
  model of type theory  by induction on type structure.} 
\end{minipage}
} \par}

\vspace{0.3cm}

The challenge in realizing Tait's program is that the equality relation,
being defined externally, is a priori valued in the types of the meta-level.
In order to talk about equality \emph{within} the system, we must
reflect this relation from the meta-level back into the syntax.  This
step is somewhat delicate, and indeed has restricted much work along
these lines to ``truncated'' systems, in which the equality type
\emph{cannot} be iterated to yield an infinite tower
\[ p,q : a \ee A b,\qquad 
\mathpzc{p}, \mathpzc{q} : p \ee {a \ee A b} q, \qquad
\pi, \xi :\mathpzc{p} \ee {p \ee {a \ee A b} q} \mathpzc{q}, \qquad
 \cdots \]

At the same time, we certainly do want to iterate the equality type,
in order for it to generalize Martin-L\"of
intensional identity type $\idtype\!_A$.

Herein lies our main contribution.  We describe a type system $\leeq$
in which
the logical relation is reflected into the type structure via a new
type constructor, the type $A \eeq B$ of \emph{type equalities}
(between types $A, B$).  The (unique) elimination rule for this type associates to
every $e : A \eeq B$ a heterogeneous \emph{dependent equality}
${\sim} e : A \to B \to \sta$.  The introduction rules for this type
assert that every type constructor preserves type equality, including
type equality itself.  The computation rules capture the logical
conditions associated to the corresponding type constructor.

We give a
complete proof of the
\emph{preservation theorem} for $\leeq$, which states that every
expression preserves the (reflected) logical relation.  In
particular, every closed term $a : A$ is related to itself, and the
type $a \ee A a$ lives in the same universe as $A$.

For detailed development, from the simply typed lambda
calculus to the system presented here, we refer the reader to
our earlier report (\cite{ett1}).

The system $\leeq$ gives a satisfactory definion of extensional
equality for \emph{closed} types.  In order for equality to really
behave like a``type constructor'', much work remains to be done.

In the future, we would like to internalize the preservation operator, so that
extensionality of terms could be witnessed internally (cf.\ \cite{intpar}).
We also want equality to satisfy the higher-dimensional analogues
of symmetry and transitivity: the \emph{Kan filling conditions}.
An ultimate benchmark of success would be to validate all
of the axioms for equality isolated by \cite[p.34]{coquand}.

From now on, we use the words ``extensional equality''
in the sense of the thesis above.

\section{Related work}

To place our paper in context, we review some recent developments in
extensional equality.
\begin{itemize}
\item 
\emph{Observational Equality Now!}, \cite{ott}

The authors give a complete treatment of the 1-dimensional theory of equality (setoid level), including symmetry, transitivity, and the relevant computation
rules.  The constructions are performed in a metatheory having
the uniqueness of identity proofs (UIP) principle.
\item 

\emph{Equality and dependent type theory}, \cite{coquand}

These slides, which had a great influence on our own investigations,
contain early ideas for computing with univalence, taking the
syntactic rather than semantic route.
\item 
\emph{Canonicity for 2-dimensional type theory}, \cite{harper}

As stated in the title, this theory is truncated at level 2.
Nevertheless, it gives a complete computational treatment
of the groupoid operations.  The authors assume propositional reflection in the metatheory.
\item 
\emph{Computational interpretation of parametricity},
\cite{intpar}

Parametricity theory is intimately connected to extensional equality.

One key difference obtains in the
treatment of universes.  In the context of parametricity, the relation on the
universe associates to each pair of types the type of binary relations
between them.  In our notation, this would appear as the rewrite rule

\[ ({\sim} \sta^*) A B \rrule (A \to B \to \sta) \]

When defining extensional equality, we want the relation on the
universe to be type equality.  This may still allow interpretation by weaker notions of
equality --- such as isomorphism or homotopy equivalence --- but should
certainly prohibit general relations between types.

Instead, the relation on the universe in $(\leeq)$ associates to types
$A,B$ the type of equalities between $A$ and $B$:
\[ ({\sim} \sta^*) A B \rrule A \eeq B \]
The type $A \eeq B$ is thought of as the type of codes for relations with
certain properties; those properties are validated by various
notions of ``equivalence of types''.

Another difference is that the preservation theorem in parametricity results
is not iterable: even when carried out in a ``reflective'' PTS, the
witnesses of parametricity are typed in a higher universe than the
original terms.

In contrast, when we stratify $\leeq$, the ``parametricity witnesses''
will actually be typed in a \emph{lower} universe than the given
terms. (This choice will be forced upon us by semantic considerations.)

On the other hand, \cite{intpar} go much further in internalization,
reflecting the preservation operator into the syntax as well.
In our case, the preservation
map is only a meta-level operation on pseudoterms.

\item 
P. Martin-L\"of, lectures given at CMU, \cite{perML}

This talk series gives a systematic treatment of the (1-dimensional) relation model.
\item
\emph{Internalization of the groupoid model}, \cite{matthiu}

A complete formalization of the 2-dimensional theory in the Coq proof assistant.
\end{itemize}

As compared to the previous results,
our contribution internalizes the
external logical relation
in a way that neither limits the resulting theory to a low
dimension, nor requires any axioms in the metatheory.

In the next section we present the system $\lext$.
Section \ref{ext-thm} gives the proof of
Tait's extensionality theorem for $\lext$.
We use this theorem to derive extensional equality for closed types
in Section \ref{s:exte}.
Afterwards, we stratify the system to make it consistent,
and give a natural set-theoretic model.

\section{$\lext$}
In this section, we describe a type theory in which
extensionality of terms is witnessed by terms in the same system.
Denoted by $\lext$, the system is an extension of $\lstar$, the ``naive''
dependent type theory, by a new type, called
\emph{type equality}.
The typing rules for this type ensure that extensionality of every term is witnessed
within the system.
\footnote{
There is some reason to believe that $\leeq$
is a minimal dependent type theory with this property,
since it is obtained from the ``canonical'' PTS $\lstar$
by closing the type structure under reflection of the standard
logical relation.  (See (\cite{ett1}) for additional commentary.)
}

The choice of $\lstar$ as the base system is
motivated by the fact that, although inconsistent, this system is by
far and away the simplest formulation of dependent type theory.
We found that postponing proper universe management
until the rules for the new type are clearly set out
simplifies the presentation considerably.

Afterwards, the standard recipe for turning an inconsistent type
theory into a consistent one by stratifying the universes may be
applied, and the proofs given earlier remain valid.
Stratification of $\lext$ will be given in Section \ref{strat}.
 
The system admits a meta-level operation
\[ (\cdot)^* : \termsof{\lext} \to \termsof{\lext} \]
which raises by one the dimension of a given term.
Using this operation, we prove a new, fully internal form of the \emph{extensionality
  theorem} from \cite{tait}.

The dependent version of the theorem requires one to
consider a certain relation on the universe of types, and for every pair of
types related by it, a new ``heterogeneous'' relation between terms of
these types.  We shall now define these concepts.

\subsubsection*{Intuitive description}
We set out by stipulating that there be a binary relation
$\eeq\ : \sta \to \sta \to \sta$ on the universe of types.
It is a new type constructor, and we call it \emph{type equality}.
For types $A,B : \sta$, the type of equalities between $A$ and $B$ is
denoted $A \eeq B$.

Every equality $e : A \eeq B$ between $A$ and $B$ induces a binary
relation ${\sim}e : A \to B \to \sta$ between $A$ and $B$.  The
${\sim}(\cdot)$-operator is the elimination rule for the type $A \eeq B$.

For $a:A$ and $b:B$, we think of the type ${\sim} e a b$ as
representing equalities between $a$ and $b$ which are
``lying over'' $e : A \eeq B$.  To articulate this
intuition, we introduce the notation
\[ a \sim_e b \df {\sim}e a b \quad \quad \comment{: \sta}\]

We add term constructors which assert that every type constructor
preserves type equality, including type equality itself.  These terms
are the constructors for the type $A \eeq B$.

For instance, the constructor
corresponding to $\eeq$ asserts that $\eeq$ preserves type
equality.  The corresponding introduction rule becomes
\[
\AXC{$ A^* : A \eeq A'$}
\AXC{$ B^* : B \eeq B'$}
\BIC{$ \seq A^* B^* : (A \eeq B) \eeq (A' \eeq B')$}
\DisplayProof\]

Finally, for every combination of an introduction rule and elimination
rule, there must be a reduction rule specifying how the two
interact.  In $\leeq$, there are four type constructors: $\Pi$,
$\Sigma$, $\eeq$, and $\sta$.  Thus, there are $4 \times 1 = 4$
reduction rules for $\eeq$.

The reduction rules capture the logical conditions corresponding
to the four
type constructors.  They will insure that the extensional equality on
every type is \emph{definitionally} equal to a type expressible with
basic constructors.
\footnote{
This is also the reason why the type $a \sim_e b$ does not require axioms stating that
it preserves equality:\\ it is not a proper type constructor, but 
reduces to more basic types according to the structure of $e$.}

For instance, extensional equality on the $\Pi$ and $\Sigma$ types
is
\begin{align*}
  f \ee {\Pi x{:}A.B(x)} f' \df &\Pi a{:}A\Pi a'{:}A\Pi a^* : a \ee A
  a'.\ \ f a \sim_{B(a^*)} f' a' \\
  (a,b) \ee {\Sigma x{:}A. B(x)} (a',b')
  \df &\Sigma a^* : a \ee {A} a'. \ \ b \sim_{B(a^*)} b'
\end{align*}

Leaving the full treatment of extensional equality to Section \ref{s:exte},
suffice it to say that the reduction rules for type equality are wholly
motivated by generalizing these ``logical conditions'' to the dependent
case.
Conversely, our definition of extensional equality will indeed
arise as the specialization of the heterogeneous
relation $\sim_e : A \to B \to \sta$ to the case when $A = B$
and when $e$ is the ``degenerate path'' $\refl{A} : A \eeq A$.

We are ready to present the system $\leeq$.

\subsubsection*{Formal description}

\begin{description}
\item[Syntax]:
\vspace{-0.5cm}
\begin{align*}
A,B,s,t,e
::= \sta &\mid x \mid \Pi x {:} A. B \mid \Sigma x{:}A. B \mid A \eeq B \mid a \sim_e b\\
&\mid \lambda x{:}A.t \mid s t \mid (s,t) \mid \pi_1 t \mid \pi_2 t\\
&\mid \sta^* \mid \Pi^* [x,x',x^*]:A.B \mid \Sigma^* [x,x',x^*]:A.B \mid
\seq e e 
\end{align*}
\item[Typing] \optarg{\text{(greyed out font demarcates implicit arguments)}}:
\begin{prooftree}
  \AXC{$$}
  \UIC{$ \Gamma \vdash \sta : \sta$}
\end{prooftree}
\begin{prooftree}
  \AXC{$\Gamma \vdash A : \sta$}
  \UIC{$\Gamma, x:A \vdash x:A$}
  \end{prooftree}
  \begin{prooftree}
  \AXC{$\Gamma \vdash M : A$}
  \AXC{$\Gamma \vdash B : \sta$}
  \BIC{$\Gamma, y:B \vdash M:A$}
\end{prooftree}
\begin{prooftree}
  \AXC{$\Gamma \vdash A : \sta$}
  \AXC{$\Gamma, x : A \vdash B : \sta$}
  \BIC{$\Gamma \vdash \Pi x{:}A.B : \sta$}
  \noLine
  \UIC{$\Gamma \vdash \Sigma x{:}A.B : \sta$}
\end{prooftree}
\begin{prooftree}
  \AXC{$\Gamma \vdash A : \sta$}
  \AXC{$\Gamma \vdash B : \sta$}
  \BIC{$\Gamma \vdash A \eeq B : \sta$}
\end{prooftree}
\begin{prooftree}
  \AXC{$\Gamma \vdash A : \sta$}
  \AXC{$\Gamma \vdash B : \sta$}
  \AXC{$\Gamma \vdash e : A \eeq B$}
  \TIC{$\Gamma \vdash {\sim}e : A \to B \to \sta$}
\end{prooftree}

\[ \fbox{$a \sim_e b \df {\sim} e a b$} \]

  \begin{prooftree}
    \AXC{$\optarg{\Gamma \vdash A : *}\quad
 \optarg{\Gamma, x: A \vdash B : *}$}
    \AXC{$\Gamma, x : A \vdash b : B$}
    \BIC{$\Gamma \vdash \lambda x{:}A.b : \Pi x{:}A. B$}
  \end{prooftree}
  \begin{prooftree}
    \AXC{$\optarg{\Gamma \vdash A : *}\quad
 \optarg{\Gamma, x: A \vdash B : *}$}
    \AXC{$\Gamma \vdash f : \Pi x{:}A.B$}
    \AXC{$\Gamma \vdash a : A$}
    \TIC{$\Gamma \vdash f a : B[a/x]$}
  \end{prooftree}
\begin{prooftree}
    \AXC{$\optarg{\Gamma \vdash A : *}\quad
 \optarg{\Gamma, x: A \vdash B : *}$}
  \AXC{$\Gamma \vdash a : A$}
  \AXC{$\Gamma \vdash b : B[a/x]$}
  \TIC{$\Gamma \vdash (a,b) : \Sigma x{:}A.B$}
\end{prooftree}
  \begin{prooftree}
    \AXC{$\optarg{\Gamma \vdash A : *}\quad
 \optarg{\Gamma, x: A \vdash B : *}$}
    \AXC{$\Gamma \vdash p : \Sigma x{:}A.B$}
    \BIC{$\Gamma \vdash \pi_1 p : A\phantom{[\pi_1 p / x]}$}
    \noLine
    \UIC{$\Gamma \vdash \pi_2 p : B[\pi_1 p / x]$}
  \end{prooftree}
\begin{prooftree}
  \AXC{$\Gamma \vdash M : A$}
  \AXC{$\Gamma \vdash B : *$}
  \AXC{$A = B$}
  \TIC{$\Gamma \vdash M : B$}
\end{prooftree}
\begin{prooftree}
\AXC{$$}
\UIC{$\Gamma \vdash \sta^* : \sta \ee{} \sta$}
\end{prooftree}
\begin{prooftree}
\AXC{$\begin{aligned}
&\optarg{\Gamma \vdash A:*}\\
&\optarg{\Gamma \vdash A':*}\\
&\phabra\Gamma \vdash A^* : A \ee{} A'
\end{aligned}$}
\AXC{$\begin{aligned}
&\optarg{\Gamma, x:A \vdash B:*}\\
&\optarg{\Gamma, x':A' \vdash B':*}\\
\Gamma, x{:}A, x'{:}A',\, &x^*: x {\sim_{A^*}} x' \vdash B^* : B \ee{} B'
\end{aligned}$}
\BIC{$\Gamma \vdash \Pi^*\, [x,x',x^*] : A^*.\, B^* : \Pi x{:}A.B \ee{} \Pi x'{:}A'.B'$}
\noLine
\UIC{$\Gamma \vdash \Sigma^*\, [x,x',x^*] : A^*.\, B^* : \Sigma x{:}A.B \ee{} \Sigma x'{:}A'.B'$}
\end{prooftree}
\begin{prooftree}
\AXC{$\begin{aligned}
&\optarg{\Gamma \vdash A:*}\\
&\optarg{\Gamma \vdash A':*}\\
&\phabra\Gamma \vdash A^* : A \ee{} A'
\end{aligned}$}
\AXC{$\begin{aligned}
&\optarg{\Gamma \vdash B:*}\\
&\optarg{\Gamma \vdash B':*}\\
&\phabra\Gamma \vdash B^* : B \ee{} B'
\end{aligned}$}
\BIC{$\Gamma \vdash \seq A^* B^* : (A \ee{} B) \ee{} (A' \ee{} B')$}
\end{prooftree}
\item[Reduction]:
\begin{align*}
(\lambda x{:}A.s)t &\rrule s[t/x]\\
\pi_i(s_1,s_2) &\rrule s_i\\
A \sim_{\sta^*} B &\rrule A \eeq B\\
f \sim_{\Pi^* [x,x',x^*] : A^*. B^*} f' &\rrule \Pi x{:}A
\Pi x'{:}A' \Pi x^* : x \sim_{A^*} x'.\ f x \sim_{B^*} f' x'\\
p \sim_{\Sigma^* [x,x',x^*] : A^*. B^*} p' &\rrule \Sigma a^* : \pi_1 p
  \sim_{A^*} \pi_1 p'.\ \pi_2 p \sim_{B^*[\pi_1 p, \pi_1 p',
  a^*/x,x',x^*]} \pi_2 p'\\
e \sim_{\seq A^* B^*} e' &\rrule 
    \Pi a {:}A \Pi a'{:}A' \Pi a^* : a
    \sim_{A^*} a'\\
    &\phantom{\rrule {}.} \Pi b{:}B \Pi b'{:}B' \Pi b^* : b \sim_{B^*}
    b'.\quad (a \sim_e b) \eeq (a' \sim_{e'} b')
  \end{align*}
\end{description}
In what follows, we will often see a pattern where three operations
of the same type appear in a row,
like the triple-product sequences in the last reduction rule.
To reduce clutter in such expressions, we introduce the following notations.

\begin{align*}
  \prod \tripar{x:A}{y:B}{z:C}\, T &\de \Pi x{:}A \Pi y{:}B \Pi z{:}C. T\\
  \blam \tripar{x:A}{y:B}{z:C}\, t &\de \lambda x{:}A \lambda
  y{:}B \lambda z {:} C. t\\
  M \tripar{N_1}{N_2}{N_3} &\de M N_1 N_2 N_3\\
  M \trip{a/x}{b/y}{c/z} &\de M[a/x][b/y][c/z]\\
  \prod\hista \trip{x}{y}{z} : A^*. B^* &\de \Pi^* [x,y,z] : A^* . B^*\\
  \sum\hista \trip{x}{y}{z} : A^*. B^* &\de \Sigma^* [x,y,z] : A^* . B^*
\end{align*}

With these conventions, the reduction rules for the type $A \eeq B$ may
be rendered as
\begin{align*}
      A \sim_{\sta^*} B &\rrule A \ee{} B\\
    f \sim_{\Pi^* [x,x',x^*] : A^*. B^*} f' &\rrule \prod \tripar{a:A}{a':A'}{a^* : a
    {\sim_{A^*}} a'}\ \ssim B^*\trip{a/x}{a'/x'}{a^*/x^*} \ f x\ f'\! x'\\
    p \sim_{\Sigma^* [x,x',x^*] : A^*. B^*} p' &\rrule \sum_{a^* : \pi_1 p {\sim_{A^*}}
    \pi_1 p'} \ssim B^*\trip{\pi_1 p/x}{ \!\!\pi_1 p'/x'}{\,\,\,a^*/x^*}\ \pi_2 p\ 
    \pi_2 p'\\
    e \sim_{\seq A^* B^*} e' &\rrule \prod \tripar{a :A}{a':A'}{a^* : a
    {\sim_{A^*}} a'} \prod \tripar{b:B}{b':B'}{b^* : b {\sim_{B^*}}
    b'}\ (a \sim_e b) \eeq (a' \sim_{e'} b') 
\end{align*}

\section{The $(\cdot)^*$-operator} \label{ext-thm}
We now define the map $(\cdot)^* : \termsof{\lext} \to
\termsof{\lext}$ which will satisfy
\[t = t(x_1,\dots, x_n) : A(\vec x)
\qquad \then \qquad t^* = t^*
{\scriptsize{\left(\begin{array}{c} x_1\\ x_1' \\x_1^* \end{array}\right.
\cdot \ \cdot \ \cdot
\left.\begin{array}{c} x_n\\ x_n' \\x_n^* \end{array}\right)}}
 : t(\vec x) \sim_{A^*(\vec x, \vec x', \vec x^*)} t' (\vec x')\]
The intuition is that $t^*$ gives the transport of $t$
over a ``formal path'' in the context.
\newpage

\begin{definition}
  Let $t \mapsto t'$ be the operation of apostrophizing every variable,
bound or otherwise.
\end{definition}

\begin{lemma}
    \begin{itemize}
    \item $(M[N/x])' = M'[N'/x']$
    \item $M=N \then M'=N'$
    \item $\Gamma \vdash
    M : A \then \Gamma' \vdash M' : A'$
    \end{itemize}
  \end{lemma}
  \begin{proof}
   Typography.  
  \end{proof}

  \begin{definition}
The operation $t \mapsto t^*$ is defined by induction on term structure.

In the equations that follow, the symbols $A_*$, $B_*$, $a_*$, etc.\ are free
variables: the appearence of $*$ in a {\bf sub}script is merely a suggestive choice of
naming the variables.

\begin{align*}
  (\sta)^* &\eq \sta^*\\
  (x)^* &\eq x^*\\
  (\Pi x{:}A. B)^* &\eq \prod\hista \trip{x}{x'}{x^*} : A^*.\ B^*\\
  (\Sigma x{:} A. B)^* &\eq \sum\hista \trip{x}{x'}{x^*} : A^*.\ B^*\\
  (A \eeq B)^* &\eq \seq A^* B^*\\
(\ssim e)^* &\eq e^*\\
(\lambda x{:}A.b)^* &\eq \lambda x{:}A\; \lambda x'{:}A'\;
  \lambda x^* : x {\sim_{A^*}} x'.\   b^*
  \\(f a)^* &\eq f^* a a' a^* 
  \\(a,b)^* &\eq (a^*,b^*)\\
  (\pi_1 p)^* &\eq \pi_1 p^*\\
  (\pi_2 p)^* &\eq \pi_2 p^*\\
 (\sta^*)^* &\eq \blam \tripar {A : \sta }{A' : \sta} {A^* : A \eeq A'}
\blam \tripar {B : \sta} {B': \sta} {B^* : B \eeq B'}.\ \seq A^* B^*
\\
 (\Pi^* [x,x_1,x_*] : A_*.\ B_*)^* &\eq
\blam \tripar { f : \Pi x{:}A.B}{ f':\Pi x'{:}A'.B'}{ f^*
 : f \sim_{\Pi^* A^* B^*} f'}
\blam \tripar {f_1 : \Pi x_1{:}A_1.B_1} {f'_1:\Pi x_1' {:}A_1'.B_1'} {f^*_1
 : f_1 \sim_{\Pi^* A^*_1 B^*_1} f'_1}.\\
&\phantom{\eq\;}\,
\prod\hista \trip{a}{a'}{a^*} : A^*\;
\prod\hista \trip{a_1}{a_1'}{a_1^*} : A_1^*\;
\prod\hista \trip{a_* }{a_*'}{a_*^*} : A_*^*  \tripar{a}{a'}{a^*} \tripar{a_1}{a'_1}{a^*_1}.\\
&\phanq
B^*_* \trip{a/x}{a'/x'}{a^*/x^*}
\trip{a_1/x_1}{a_1'/x_1'}{a_1^*/x_1^*}
\trip{a_*/x_*}{a_*'/x_*'}{a_*^*/x_*^*}
\tripar{f a}{f' a'}{f^* a a' a^*}
\tripar{f_1 a_1}{f'_1 a'_1}{f^*_1 a_1 a'_1 a^*_1}
\\
 (\Sigma^* [x,x_1,x_*] : A_*.\ B_*)^*
&\eq \blam \tripar{p : \Sigma x{:}A.B}{p':\Sigma x'{:}A'.B'}
{p^* : p \sim_{\Sigma^* A^* B^*} p'}
\blam \tripar {p_1 : \Sigma x_1{:}A_1.B_1} {p_1' : \Sigma x_1'{:}A_1'.B_1'} {p_1^* : p_1 \sim_{\Sigma^* A_1^* B_1^*} p_1'}.\\
&\phan
\sum\hista\trip{a_* : \pi_1 p \sim_{A_*} \pi_1 p_1}
{a'_* : \pi_1 p' \sim_{A'_*}  \pi_1 p'_1}
{a^*_* : a_* \sim_{(A^\sim_*\, \pi\!{}_1\! p\; \pi\!{}_1\! p\!{}_1)^*} a'_*} :
A^*_* \tripar{\pi_1 p}{\pi_1 p'}{\pi_1 p^*}
\tripar{\pi_1 p_1} {\pi_1 p_1'} {\pi_1 p_1^*}.\\
&\phanq B^*_* \trip{\pi_1 p/x}{\pi_1 p'/x'}{\pi_1 p^*/x^*}
\trip{\pi_1 p_1/x_1}{\pi_1 p_1'/x_1'}{\pi_1 p_1^*/x_1^*}
\trip{a_*/x_*}{a_*'/x_*'}{a_*^*/x_*^*}
\tripar{\pi_2 p}{\pi_2 p'}{\pi_2 p^*}
\tripar{\pi_2 p_1}{\pi_2 p_1'}{\pi_2 p_1^*}
\end{align*}
\begin{align*}
 (\seq A_* B_*)^* &\eq \blam \tripar{e : A \eeq B}
{e':A' \eeq B'} {e^* : e \sim_{\seq A^* B^*} e'}
\quad\blam\tripar{e_1 : A_1 \eeq B_1}{e_1' : A_1' \eeq B_1'}
{e_1^* : e_1 \sim_{\seq A_1^* B_1^*} e_1'}.\\
&\phanq
\prod\hista \trip{a}{a'}{a^*} : A^*\;
\prod\hista \trip{a_1}{a_1'}{a_1^*} : A_1^*\;
\prod\hista \trip{a_*}{a_*'}{a_*^*} : A_*^* \tripar{a}{a'}{a^*} \tripar{a_1}{a_1'}{a_1^*}\\
&\phanq
\prod\hista \trip{b}{b'}{b^*} : B^*\;
\prod\hista \trip{b_1}{b_1'}{b_1^*} : B_1^*\;
\prod\hista \trip{b_*}{b_*'}{b_*^*} : B_*^* \tripar{b}{b'}{b^*} \tripar{b_1}{b_1'}{b_1^*}.\\
&\phanq\qquad \bseq \left(e^* \tripar{a}{a'}{a^*}
  \tripar{b}{b'}{b^*}\right)
\left(e_1^* \tripar{a_1}{a_1'}{a_1^*} \tripar{b_1}{b_1'}{b_1^*}\right)
\vspace{-1cm}
\end{align*}
\end{definition}

\begin{lemma} \label{subst}
$(M[N/x])^* = M^*[N/x,N'/x',N^*/x^*]$
\end{lemma}

\begin{proof}
  By induction on the structure of $M$.
\end{proof}

\begin{lemma} \label{conv}
$M =N \then M^* = N^*$
\end{lemma}

\begin{proof}
By induction on the length of the reduction--expansion sequence in
  $M=N$, it suffices to show
  \begin{align} \label{conv}
    M \to N \;\then\; M^* \thra N^*
  \end{align}

First we argue that it is enough to consider contractions at the root
of the term.

Indeed, suppose that $M = C[s]$, $N = C[t]$, and $s \to t$ by
contraction
at the root.

Let $C_0 = C[x_0]$, where $x_0$ is fresh.

Using Lemma \ref{subst}, we write
\begin{align*}
  M^* &= C_0[s/x_0]^* =C_0^*[s/x_0,s'/x'_0,s^*/x^*_0]\\
  N^* &= C_0[t/x_0]^* = C_0^*[t/x_0,t'/x'_0,t^*/x^*_0]
\end{align*}
Since $\beta$-reduction is itself a congruence, it suffices to verify
that the instances of each variable are reducible.  That $s \thra t$ and $s' \thra
t'$
is clear; that $s^* \thra t^*$ remains to be proved.

There is thus no loss of generality in assuming that the
redex is contracted at the root.

We now treat each reduction rule in order.

\newcommand{\st}{.\hspace{0.5em}}
\begin{description}
\item[\underline{$\beta$}] Given $s = (\lambda x : A. M) N \to M[N/x]
  = t$, we are to show that
\[((\lambda x {:} A. M) N)^* \thra M[N/x]^*\]
Indeed,
\begin{align*}
  ((\lambda x {:} A. M) N)^* &= (\lambda x{:}A.M)^* N N' N^*\\
&= (\lambda x{:}A \lambda x'{:}A' \lambda x^* : x \sim_{A^*} x'. M^*)
N N' N^*\\
&\thra M^*[N/x,N'/x',N^*/x^*]\\
&= M[N/x]^*
\end{align*}
by Lemma \ref{subst}.  Thus $s^* \thra t^*$.
\item[\underline{$\beta_\Sigma$}] We have
  \begin{align*}
    (\pi_1 (M,N))^* = \pi_1 (M,N)^* &= \pi_1 (M^*,N^*) \to M^*\\
    (\pi_2 (M,N))^* = \pi_2 (M,N)^* &= \pi_2 (M^*,N^*) \to N^*
  \end{align*}
\item[\underline{$\sta^*$}] Consider \[\ssim \sta^* \too \lambda A{:}\sta \lambda
  B{:}\sta.\; A \eeq B\]
We have 
\begin{align*} (\ssim \sta^*)^*
&\eq \lambda A{:}\sta \lambda A'{:}\sta \lambda A^* : A \sim_{\sta^*} A'\\
&\qquad\quad\lambda B{:}\sta \lambda B'{:}\sta \lambda B^* : B \sim_{\sta^*} B'.\;
\seq A^* B^*\\
&\eq (\lambda A{:}\sta \lambda B{:}\sta{}.\; A \eeq B)^*
\end{align*}
\item[\ul{$\Pi^\sta$}]  Consider
\begin{align*}
&\ssim (\Pi^* [x,x_1,x_*]:A_*. B_*)\\ &\too \lambda f : \Pi x{:}A. B \; \lambda f_1 :
\Pi x_1{:}A_1.B_1.\\
&\phanq\quad  \Pi a{:}A \Pi a_1{:}A_1 \Pi a_* : a \sim_{A_*} a_1
\st f a \sim_{B_* [a a_1 a_*/x x_1 x_*]} f_1 a_1
\end{align*}
Let $T$ be the reduct on the right.  We have
\begin{align*}
  &(\sim \Pi^*[x,x_1,x_*] : A_* B_*)^* \\ &\eq \blam \trip{f : \Pi x{:}A.B}
{f' : \Pi x'{:}A'. B'}{f^* : f \sim_{\Pi^* A^* B^*} f'}\quad
\blam \trip{f_1 : \Pi x_1{:}A_1.B_1}
{f'_1 : \Pi x'_1{:}A'_1. B'_1}{f_1^* : f_1 \sim_{\Pi^* A_1^*  B^*_1}
  f'_1}.\\
&\phan
\prod\hista \trip{a}{a'}{a^*} : A^*\;
\prod\hista \trip{a_1}{a_1'}{a_1^*} : A_1^*\;
\prod\hista \trip{a_*}{a_*'}{a_*^*} :
A_*^*  \tripar{a}{a'}{a^*} \tripar{a_1}{a'_1}{a^*_1}.\\
&\phanq
B^*_* \trip{a/x}{a'/x'}{a^*/x^*}
\trip{a_1/x_1}{a_1'/x_1'}{a_1^*/x_1^*}
\trip{a_*/x_*}{a_*'/x_*'}{a_*^*/x_*^*}
\tripar{f a}{f' a'}{f^* a a' a^*}
\tripar{f_1 a_1}{f'_1 a'_1}{f^*_1 a_1 a'_1 a^*_1}
\end{align*}
By inspection, this is exactly $T^*$.  Let's check the innermost
quantifier:
\begin{align*}
&(\Pi a_* : a \sim_{A_*} a_1 \st f a \sim_{B_* [a a_1 a_*/x x_1 x_*]}
  f_1 a_1)^*\\
&\eq \prod\hista \trip{a_* : a \sim_{A_*} a_1}
{a_*' : a' \sim_{A'_*} a'_1}
{a_*^* : a_* \sim_{A^*_*} a'_*} :
A^*_* \tripar{a}{a'}{a^*} \tripar{a_1}{a_1'}{a_1^*}.\ 
(\sim B_*[a a_1 a_*/x x_1 x_*] (f a) (f_1 a_1))^*
  \end{align*}
(Here we used that $(a \sim_{A_*} a_1)^* = A_*^* a a' a^* a_1 a_1'
a_1^*$.)
Indeed,
\begin{align*}
&(\sim B_*[a a_1 a_*/x x_1 x_*] (f a) (f_1 a_1))^*\\
&\eq (B_*[a/x][a_1/x_1][a_*/x_*])^* \tripar{f a}{f' a'}{f^* a a' a^*}
\tripar{f_1 a_1}{f_1' a_1'}{f_1^* a_1 a_1' a_1^*}\\
&\eq B_*^*\trip{a/x}{a'/x'}{a^*/x^*}
\trip{a_1/x_1}{a_1'/x_1'}{a_1^*/x_1^*}
\trip{a_*/x_*}{a_*'/x_*'}{a_*^*/x_*^*}
\tripar{f a}{f' a'}{f^* a a' a^*}
\tripar{f_1 a_1}{f'_1 a'_1}{f^*_1 a_1 a'_1 a^*_1}
\end{align*}
as required.
\end{description}
Other cases are treated similarly.
\end{proof}

\begin{definition}
  Let $\Gamma = (x_1 : A_1, \dots, x_n : A_n)$
be a context.  Put
\[ \Gamma^* \df \left(
  \begin{array}{cccc}
    x_1 : A_1 & \phantom{blue} && x_n : A_n\\
    x_1' : A_1' &\cdots& \cdots& x'_n : A'_n\\
    x_1^* : x_1 \sim_{A_1^*} x_1' &&\phantom{blue}& x_n^* : x_n \sim_{A_n^*} x_n'
  \end{array}\right)\]
\end{definition}

\newpage
\begin{theorem}
\qquad\qquad$\Gamma \vdash M : A \quad \then \quad \Gamma^* \vdash M^* : M
\sim_{A^*} M'$
\end{theorem}

\begin{proof}
  We proceed by induction on $\Gamma \vdash M : A$.
  \begin{description}
  \item[Axiom] For the axiom rule $\overline{\vdash \sta : \sta}$, we have
\[ \vdash \sta^* : \sta \eeq \sta \]
By conversion rule, $\vdash \sta^* : \sta \sim_{\sta^*} \sta$.
\item[Variable] Given the derivation
\begin{prooftree}
  \AXC{$\Gamma \vdash A : \sta$}
  \UIC{$\Gamma, x:A \vdash x:A$}
  \end{prooftree}
we have, by induction hypothesis, that
\[ \Gamma^* \vdash A^* : A \sim_{\sta^*} A' \]
and hence $A^* : A \eeq A'$.

Clearly, $\Gamma^* \supseteq \Gamma' \vdash A' : \sta$.

Then $(\Gamma, x:A)^* = (\Gamma^*, x{:}A, x'{:}A', x^* : x \sim_{A^*}
x')$ is a valid context, and
\[ (\Gamma, x:A)^* \vdash x^* : x \sim_{A^*} x'\]
\item[Weakening] Given the derivation
  \begin{prooftree}
  \AXC{$\Gamma \vdash M : A$}
  \AXC{$\Gamma \vdash B : \sta$}
  \BIC{$\Gamma, y:B \vdash M:A$}
\end{prooftree}
the induction hypothesis gives that
\begin{align*}
  \Gamma^* &\vdash M^* : M \sim_{A^*} M'\\
  \Gamma^* &\vdash B^* : B \eeq B'
\end{align*}
Then $(\Gamma, y:B)^* = (\Gamma^*, y{:}B, y'{:}B', y^* : y
\sim_{B^*} y')$ is a valid context, and
\[ (\Gamma, y:B)^* \vdash M^* : M \sim_{A^*} M'\]
(by applying weakening thrice).

\item[Formation of $\Pi$,$\Sigma$] 
Suppose we are given
\begin{prooftree}
  \AXC{$\Gamma \vdash A : \sta$}
  \AXC{$\Gamma, x : A \vdash B : \sta$}
  \BIC{$\Gamma \vdash \Pi x{:}A.B : \sta$}
\end{prooftree}
Induction gives
\begin{align*}
  \Gamma^* &\vdash A^* : A \eeq A'\\
  \Gamma^*, x{:}A, x'{:}A', x^* : x \sim_{A^*} x'
&\vdash B^* : B \eeq B'
\end{align*}
By lemmata, we also have apostrophized versions of these:
\begin{align*}
  \Gamma' \vdash A' : \sta\\
  \Gamma', x' : A' \vdash B' : \sta 
\end{align*}
Together, the given data, the primed version, and the inductive
version,
provide the hypotheses necessary for the application of the $\Pi^*$-rule:
\begin{prooftree}
  \AXC{$\begin{aligned}
&\Gamma \vdash A:*\\
&\Gamma \vdash A':*\\
&\Gamma \vdash A^* : A \ee{} A'
\end{aligned}$}
\AXC{$\begin{aligned}
\Gamma, x:A &\vdash B:*\\
\Gamma, x':A' &\vdash B':*\\
\Gamma, x:A, x':A', x^*: x \sim_{A^*} x' &\vdash B^* : B \ee{} B'
\end{aligned}$}
\BIC{$\Pi^*\, [x,x',x^*] : A^*.\, B^* : \Pi x{:}A.B \ee{} \Pi x'{:}A'.B'$}
\end{prooftree}
Since $(\Pi x{:} A. B)^* = \Pi^* [x,x',x^*] : A^*. B^*$, the above
judgement has the desired form.

The case of $\Sigma$-formation is treated congruently.

\item[$\eeq$-Formation]  This is like the previous case, but easier; given
\begin{prooftree}
  \AXC{$\Gamma \vdash A : \sta$}
  \AXC{$\Gamma \vdash B : \sta$}
  \BIC{$\Gamma \vdash A \eeq B : \sta$}
\end{prooftree}
we have that $\Gamma' \vdash A' : \sta, \Gamma' \vdash B' : \sta$, and also, by induction, that
\begin{align*}
  \Gamma^* &\vdash A^*: A \eeq A'\\
  \Gamma^* &\vdash B^* : B \eeq B'
\end{align*}
These data allow us to apply the $\seq$-rule, yielding
\[\Gamma^* \vdash \seq A^* B^* : (A \eeq B) \eeq (A' \eeq B') \]
which type converts to $(A \eeq B) \sim_{\sta^*} (A \eeq B)'$, as required.
\item[${\sim}$-Formation]
Suppose we are given
\begin{prooftree}
  \AXC{$\Gamma \vdash A : \sta$}
  \AXC{$\Gamma \vdash B : \sta$}
  \AXC{$\Gamma \vdash e : A \eeq B$}
  \TIC{$\Gamma \vdash {\sim} e : A \to B \to \sta$}
\end{prooftree}
Then we have $\Gamma' \vdash {\sim} e' : A' \to B' \to \sta$, and by
induction
\[\Gamma^* \vdash e^* : e \sim_{(A \eeq B)^*} e'\]

We work in $\Gamma^*$.  Reducing the type of $e^*$, we get
\begin{align*}
e^* \of &\prod \tripar{a:A}{a':A'}{a^* : a {\sim_{A^*}} a'}
\prod \tripar{b:B}{b':B'}{b^*: b {\sim_{B^*}} b'}\quad (a \sim_e b)
\eeq (a' \sim_{e'} b')
\end{align*}
When written in explicit form, this looks like
\[ e^* : \prod \tripar{a}{a'}{a^*} \prod \tripar{b}{b'}{b^*}\quad ({\sim} e a b) \eeq
({\sim} e' a' b') \]

By definition of $(\cdot)^*$, 
\[(A \to B \to \sta)^* = \Pi^* [x,x',x^*] : A^*\; \Pi^* [y,y',y^*] :
B^*. \sta^*\]

Thus, for any $E : A \to B \to \sta$, $E' : A' \to B' \to \sta$, we
find that
\begin{align*}
E \sim_{(A \to B \to \sta)^*} E' \eq
&\prod \tripar{x:A}{x':A'}{x^* : x
\sim_{A^*} x'} \prod \tripar { y:B }{y':B'}{y^*:y \sim_{B^*} y'}\;E x
y \eeq E' x' y'
\end{align*}

In particular, the type of $e^*$ is exactly 
\[( {\sim} e) \sim_{(A \to B \to \sta)^*} ({\sim} e') \]

Since $(\ssim e)^* = e^*$, we conclude that
\[ \Gamma^* \vdash (\ssim e)^* : (\ssim e) \sim_{(A \to B \to \sta)^*}
(\ssim e)'\]
\item[Abstraction] Given
  \begin{prooftree}
    \AXC{$\Gamma \vdash A : *$}
    \AXC{$\Gamma, x : A \vdash B : *$}
    \AXC{$\Gamma, x : A \vdash b : B$}
    \TIC{$\Gamma \vdash \lambda x{:}A.b : \Pi x{:}A. B$}
  \end{prooftree}
We have, by IH, that
\begin{align*}
  \Gamma^* &\vdash A^* : A \eeq A'\\
  \Gamma^*, x{:}A, x'{:}A', x^* : x{\sim_{A^*}}x' &\vdash B^* : B \eeq
  B'\\
  \Gamma^*, x{:}A, x'{:}A', x^* : x{\sim_{A^*}}x' &\vdash b^* : b
  \sim_{B^*} b'
\end{align*}
Observe that our target type converts as
\begin{align*}
&\quad (\lambda x{:}A.b) \sim_{(\Pi x{:}A. B)^*} (\lambda x'{:}A'.b')\\
&= \prod \tripar{a:A}{a':A'}{a^* : a {\sim_{A^*}} a'}\
(\lambda x{:}A.b) a \sim_{B^* [a,a',a^*/x,x',x^*]}
(\lambda x'{:}A'.b') a'\\
&= \prod \tripar{a:A}{a':A'}{a^* : a {\sim_{A^*}} a'}\
                b[a/x] \sim_{B^*[a,a',a^*/x,x',x^*]} b'[a'/x']\\
&=_\alpha \prod \tripar{x:A}{x':A'}{x^* : x {\sim_{A^*}} x'}\
                b \sim_{B^*} b'
              \end{align*}
The first two induction hypotheses give us that this is a
well-formed type.  The third, after three applications of
the abstraction rule, gives
\[\Gamma^* \vdash (\lambda x{:}A \lambda x'{:}A' \lambda x^* : x {\sim_{A^*}} x'.\
                b^*) : (\Pi x{:}A \Pi x'{:}A' \Pi x^* : x {\sim_{A^*}} x'.\
                b \sim_{B^*} b')\]
The subject of this judgement is equal to $(\lambda x{:}A.b)^*$.\\
The type predicate converts to $(\lambda x{:}A.b) \sim_{(\Pi x{:}A. B)^*} (\lambda x'{:}A'.b')$.
\item[Application] Suppose we are given
  \begin{prooftree}
    \AXC{$\Gamma \vdash A : *\quad \Gamma, x: A \vdash B : *$}
    \AXC{$\Gamma \vdash f : \Pi x{:}A.B$}
    \AXC{$\Gamma \vdash a : A$}
    \TIC{$\Gamma \vdash f a : B[a/x]$}
  \end{prooftree}
The induction hypotheses are
\begin{align}
\notag \Gamma^* &\vdash A^* : A \eeq A'\\
\notag \Gamma^*, x{:}A, x'{:}A', x^* : x {\sim_{A^*}} x' &\vdash B^* : B \eeq B'\\
\Gamma^* &\vdash f^* : f \sim_{(\Pi x{:}A.B)^*} f' \label{app3}\\
\notag \Gamma^* &\vdash a^* : a \sim_{A^*} a'
\end{align}
Working in $\Gamma^*$, we need to show that
\[(fa)^* : fa \sim_{B[a/x]^*} f'a'\]
Equivalently, we need to show that
\begin{align}
  f^* a a' a^* : &fa \sim_{B[a,a',a^*/x,x',x^*]} f'a' \label{appGoal}
\end{align}
(where we used the substitution lemma to rewrite $B[a/x]^*$).

Applying the converison rule to \eqref{app3} gives
\[f^* : = \prod \tripar{x:A}{x':A'}{x^* : x {\sim_{A^*}} x'} \ fx \sim_{B^*}
f'x'\]
Then, by a triple use of the application rule, we have
\[f^* a a' a^* : f a \sim_{B^*[a,a',a^*/x,x',x^*]} f' a' \]
which is typographically consistent with \eqref{appGoal}.
\item[Pairing] Let us be given
\begin{prooftree}
  \AXC{$\Gamma \vdash A : * \quad \Gamma, x:A \vdash B : *$}
  \AXC{$\Gamma \vdash a : A$}
  \AXC{$\Gamma \vdash b : B[a/x]$}
  \TIC{$\Gamma \vdash (a,b) : \Sigma x{:}A.B$}
\end{prooftree}
We work in $\Gamma^*$.  By induction, we have
\begin{align*}
  a^* &: a \sim_{A^*} a'\\
  b^* &: b \sim_{B[a/x]^*} b'
\end{align*}
We may rewrite the latter as
\begin{align}
  \label{Btyp}
b^* :  b \sim_{B^*[a,a',a^*/x,x',x^*]} b'
\end{align}

Using these data, the following sequence of judgements may be verified:
\begin{align*}
 (\Gamma, x:A)^* &\vdash B^* : B \eeq B'\\
 \Gamma^*, x^* : a \sim_{A^*} a' &\vdash B^*[a,a',x^*/x,x',x^*] :
 B[a/x] \eeq B'[a'/x']\\
\Gamma^*, x^* : a \sim_{A^*} a' &\vdash b
\sim_{B^*[a,a',x^*/x,x',x^*]} b' : \sta\\
\Gamma^* &\vdash b
\sim_{B^*[a,a',a^*/x,x',x^*]} b' : \sta\\
\Gamma^* &\vdash (a^*,b^*) : \Sigma a^* : a \sim_{A^*} a'.\
b \sim_{B^*[a,a',a^*/x,x',x^*]} b'\\
\Gamma^* &\vdash (a^*,b^*) : (a,b) \sim_{\Sigma^* [x,x',x^*] : A^*. B^*} (a',b')\\
\Gamma^* &\vdash (a,b)^* : (a,b) \sim_{(\Sigma x{:}A B)^*} (a',b')
\end{align*}
\item[Projections] Next, we consider the inference rules
  \begin{prooftree}
    \AXC{$\Gamma \vdash A : \sta$}
    \AXC{$\Gamma, x:A \vdash B : \sta$}
    \AXC{$\Gamma \vdash p : \Sigma x{:}A.B$}
    \TIC{$\Gamma \vdash \pi_1 p : A$}
    \noLine
    \UIC{$\Gamma \vdash \pi_2 p : B[\pi_1 p / x]$}
  \end{prooftree}
We have
\begin{align*}
  \Gamma^* &\vdash p^* : p \sim_{(\Sigma x{:}A.B)^*} p'\\
  \Gamma^* &\vdash p^* : p \sim_{\Sigma [x,x',x^*] : A^*. B^*} p'\\
  \Gamma^* &\vdash p^* : \Sigma a^* : \pi_1 p \sim_{A^*} \pi_1 p'.
  \pi_2 p \sim_{B^*[\pi_1 p,\pi_1 p', a^*/x,x',x^*]} \pi_2 p'\\
  \Gamma^* &\vdash \pi_1 p^* : \pi_1 p \sim_{A^*} \pi_1 p' \\
  \Gamma^* &\vdash (\pi_1 p)^* : (\pi_1 p) \sim_{A^*} (\pi_1 p)'
  \tag{p1}\\
  \Gamma^* &\vdash \pi_2 p^* : \pi_2 p \sim_{B^*[\pi_1 p, \pi_1 p',
    \pi_1 p^*/x,x',x^*]} \pi_2 p'\\
  \Gamma^* &\vdash (\pi_2 p)^* : (\pi_2 p) \sim_{B^*[\pi_1 p, (\pi_1 p)',
    (\pi_1 p)^*/x,x',x^*]} (\pi_2 p)'\\
   \tag{p2}  
 \Gamma^* &\vdash (\pi_2 p)^* : (\pi_2 p) \sim_{B[\pi_1 p/x]^*} (\pi_2
  p)' 
\end{align*}
The judgements (p1) and (p2) are of the required form.
\item[Conversion] Next, suppose we are given the inference
\begin{prooftree}
  \AXC{$\Gamma \vdash M : A$}
  \AXC{$\Gamma \vdash B : *$}
  \AXC{$A = B$}
  \TIC{$\Gamma \vdash M : B$}
\end{prooftree}
We are to show that $\Gamma^* \vdash M^* : M \sim_{B^*} M'$.

From the given data, we know that
\begin{align*}
  &M : B\\
  &M' : B'\\
  &B^* : B \eeq B'
\end{align*}
Thus \begin{equation}
\label{M-conv}
M \sim_{B^*} M' : \sta
\end{equation}

By Lemma \ref{conv}, we have
\begin{equation} \label{B-conv} M \sim_{A^*} M' = M \sim_{B^*} M'
\end{equation}

By IH, we also have $\Gamma^* \vdash M^* : M \sim_{A^*} M'$.
Using \eqref{M-conv} and \eqref{B-conv},
we may apply the conversion rule to this judgment
to obtain
\[
  \Gamma^* \vdash M^* : M \sim_{B^*} M'
\]
\item[$\sta$-Congruence]  Let us be given
\begin{prooftree}
\AXC{$$}
\UIC{$\sta^* : \sta \ee{} \sta$}
\end{prooftree}
We are asked to show that
\begin{align*}
  (\sta^*)^* \quad:\quad &\sta^* \sim_{(\sta \eeq \sta)^*} \sta^*\\
  (\sta^*)^* \quad:\quad &\sta^* \sim_{\seq \sta^* \sta^*} \sta^*\\
 (\sta^*)^* \quad : \quad &\prod \tripar{A :\sta}{A':\sta}{A^* : A {\sim_{\sta^*}} A'}
\prod \tripar{ B:\sta}{ B':\sta}{ B^* : B  {\sim_{\sta^*}} B'}.\ (A \sim_{\sta^*} B)
\eeq (A' \sim_{\sta^*} B')
\end{align*}
 Unfolding the definition of $(\sta^*)^*$, we have
\[(\sta^*)^* = \blam \tripar{A:\sta}{A':\sta}{A^* : A{\eeq}A'}\;
\blam \tripar{B:\sta}{B':\sta}{B^* : B {\eeq} B'}.\; \seq A^* B^*\]
By inspection, this term has the desired type.
\item[$\Pi$-congruence] Let us be given
\begin{prooftree}
\AXC{$\begin{aligned}
\Gamma &\vdash A:*\\
\Gamma &\vdash A_1:*\\
\Gamma &\vdash A_* : A \ee{} A_1
\end{aligned}$}
\AXC{$\begin{aligned}
\Gamma, x:A &\vdash B:*\\
\Gamma, x_1:A_1 &\vdash B_1:*\\
\Gamma, x:A, x_1:A_1, x_*: x \sim_{A_*} x_1 &\vdash B_* : B \ee{} B_1
\end{aligned}$}
\BIC{$\Pi^*\, [x,x_1,x_*] : A_*.\, B_* : \Pi x{:}A.B \ee{} \Pi x_1{:}A_1.B_1$}
\end{prooftree}
Recall that
\begin{align}
& (\Pi^* [x,x_1,x_*] : A_*.\ B_*)^* \label{picongterm}\\ &\eq
\blam \tripar { f : \Pi x{:}A.B}{ f':\Pi x'{:}A'.B'}{ f^*
 : f \sim_{\Pi^* A^* B^*} f'}\notag
\blam \tripar {f_1 : \Pi x_1{:}A_1.B_1} {f'_1:\Pi x_1' {:}A_1'.B_1'} {f^*_1
 : f_1 \sim_{\Pi^* A^*_1 B^*_1} f'_1}.\\
&\phantom{\eq\;}\,
\prod\hista \trip{a}{a'}{a^*} : A^*\;
\prod\hista \trip{a_1}{a_1'}{a_1^*} : A_1^*\;\notag
\prod\hista \trip{a_* }{a_*'}{a_*^*} : A_*^*  \tripar{a}{a'}{a^*} \tripar{a_1}{a'_1}{a^*_1}.\\
&\phanq
B^*_* \trip{a/x}{a'/x'}{a^*/x^*}
\trip{a_1/x_1}{a_1'/x_1'}{a_1^*/x_1^*}
\trip{a_*/x_*}{a_*'/x_*'}{a_*^*/x_*^*}
\tripar{f a}{f' a'}{f^* a a' a^*}
\tripar{f_1 a_1}{f'_1 a'_1}{f^*_1 a_1 a'_1 a^*_1} \notag
\end{align}

We are to show that
this term has type
\begin{align}
\label{pi-cong}
\Pi^* A_* B_* \sim_{\seq (\Pi x{:}A.B)^* (\Pi x_1{:}A_1.B_1)^*}
\Pi^* A'_* B'_*
\end{align}

First, note that the type of the equivalence appearing in
the index of the dependent relation in
\eqref{pi-cong} is
\[(\Pi x{:}A.B \eeq \Pi x{:}A_1.B_1) \eeq (\Pi x{:}A'.B' \eeq
\Pi x{:}A_1'.B_1')\]
(This is by the induction hypothesis on $\Pi x{:}A.B$ and $\Pi x_1{:}A_1.B_1$.
Although ``$\Pi x{:} A.B : \sta$'' does not appear among the premises of
this rule, the required statement can be obtained by inlining the proof
of the $\Pi$-formation case.  The hypotheses there are provided by
the induction hypotheses on the premises given here.)

 We begin by looking closer at the relation associated to this
 equivalence.
By $\sim$-reduction, we have
\begin{align*}
&\qquad \quad \ssim(\seq (\Pi x{:}A.B)^* (\Pi x_1{:}A_1.B_1)^*)\\
&\eq \lambda e : (\Pi x{:}A.B \eeq \Pi x_1{:}A_1.B_1)\,
  \lambda e' : (\Pi x'{:}A'.B' \eeq \Pi x_1' {:} A_1'. B_1').\\
&\phanq  \prod \tripar { f : \Pi x{:}A.B} { f' : \Pi x'{:}A'.B' }
{ f^* : f
\sim_{(\Pi x{:}A.B)^*} f'}  \prod \tripar { f_1 : \Pi x_1{:}A_1.B_1} { f_1' : \Pi
  x_1'{:}A_1'.B_1' }  { f_1^* : f_1 \sim_{(\Pi x_1{:}A_1.B_1)^*} f_1'}\\
&\phanq \qquad \qquad (f \sim_e f_1) \eeq (f' \sim_{e'} f'_1)
\end{align*}

When this term is applied to ${\Pi^* A_* B_*}$
and ${\Pi^* A_*' B_*'}$, so as to become \eqref{pi-cong},
we see immediately that $\lambda$-abstractions appearing at the root of
\eqref{picongterm} match correctly the $\Pi$-types above.

We are thus left to verify that the matrix of these abstractions 
 ---  the triple-$\Pi^*$ subexpression of \eqref{picongterm} ---  has type
\begin{align}
(f \sim_{\Pi^* A_* B_*} f_1) \eeq (f' \sim_{\Pi^* A_*' B_*'} f'_1) =
\hspace{-4cm}
\label{toprovepicong}\\
& \Pi a{:}A \Pi a_1{:}A_1 \Pi a_* : a \sim_{A_*} a_1. (f a
\sim_{B_*[a,a_1,a_*/x,x_1,x_*]} f_1 a_1)\notag\\
&   \eeq  \Pi a'{:}A' \Pi a'_1{:}A'_1 \Pi a'_* : a' \sim_{A'_*} a'_1. (f' a'
\sim_{B'_*[a',a'_1,a'_*/x',x'_1,x'_*]} f'_1 a'_1)\notag
\end{align}

Looking at \eqref{picongterm} again, we see that the
equivalences specified in the triple-$\Pi^*$ expression
correctly match the domains of the products above.
\footnote{
We recall that in order to construct an equivalence between two $\Pi$-types, one
needs to construct an equivalence between their domains of
quantification,
and, for every dependent line between these domains (ie, a
pair of terms related by the equivalence), an equivalence between
the corresponding fibers of the dependent type (quantification
matrices).

This possibility is precisely the content of the $\Pi^*$-constructor, which
gives an equivalence between two $\Pi$-terms from an equivalence between their
domains and a map transporting paths between domains to equivalences
of fibers.}
That the third equivalence
\[A^*_* \tripar{a} {a'} {a^*} \tripar{a_1} {a_1'} {a_1^*} : (a \sim_{A_*} a_1) \eeq (a'
\sim_{A'_*} a'_1) \]
has the right type uses induction hypothesis on $A_*$.

By induction hypothesis on $B_*$, we have that
\[(\Gamma, x : A, x_1:A_1,x_*: x \sim_{A_*} x_1)^*
\vdash B_*^* : B_* \sim_{\seq B^* B_1^*} B_*'\]
or, using alternative notation,
\[(\Gamma, x : A, x_1:A_1,x_*: x \sim_{A_*} x_1)^*
\vdash B_*^* : {(\seq B^* B_1^*)^\sim}B_*  B_*'\]
Plugging in the terms in our context, we get
\begin{align}
\Gamma^* &\vdash  B_*^*\trip{a/x}{a'/x'}{a^*/x^*}
\trip{a_1/x_1}{a'_1/x'_1}{a^*_1/x^*_1}  \label{bstartype} 
\trip{a_*/x_*}{a'_*/x'_*}{a_*^*/x_*^*} \\
&\of \notag
{\left(\seq B^*\trip{a/x}{a'/x'}{a^*/x^*}
B_1^*\trip{a_1/x_1}{a'_1/x'_1}{a^*_1/x^*_1}\right)}^{\bsim}
\!\!\!B_*[\vec a/ \vec x] \;
B'_*[\vec a'/\vec x']\\
&\eq \notag
\prod \tripar {y : B[a]} {y' : B'[a']} {y^*:y{\sim_{B^*[a,a',a^*]}}y'}
\prod \tripar {y_1 :B_1[a_1]} { y_1':B_1'[a_1']}
{y_1^* : y_1 \sim_{B_1^*[a_1,a_1',a_1^*]} y_1'}.
\\&\phanq\qquad  (y \sim_{B_*[\vec a]} y_1) \eeq (y' \sim_{B'_*[\vec
  a']} y_1') \notag
\end{align}
Next, we recall that
\begin{align*}
f^* &: \prod \tripar {x:A}  {x':A'} { x^* : x {\sim_{A^*}} x'}\quad f x \sim_{B^*}
f' x'\\
f^*_1 &: \prod \tripar {x_1 : A_1} {x'_1 :A_1'} {x^*: x_1 {\sim_{A^*_1}}
x'_1} \quad f_1 x_1 \sim_{B_1^*} f'_1 x'_1
\end{align*}
It folows that
\begin{align*}
  f^* a a' a^* &: {B^* \trip{a/x}{a'/x'}{a^*/x^*}}^\bsim (f a) (f' a')\\
  f_1^* a_1 a_1' a_1^* &: {B_1^* \trip{a_1/x_1}{a_1'/x_1'}{a_1^*/x_1^*}}^\bsim (f_1 a_1) (f_1' a_1')
\end{align*}
Putting this together with \eqref{bstartype}, we get
\[B^*_*
\trip{a}{a'}{a^*}\trip{a_1}{a'_1}{a^*_1}\trip{a_*}{a_*'}{a_*^*}
\tripar{fa} {f'a'}{f^* a a' a^*} \tripar {f_1 a_1} {f_1' a_1'} {f_1^* a_1 a_1'
a_1^*}\]
\[ \of (f a \sim_{B_*[\vec a]} f_1 a_1) \eeq (f' a' \sim_{B_*'[\vec
  a']} f_1' a_1')\]
This matches the expression in \eqref{toprovepicong}, concluding this case.
\item[$\Sigma$-congruence] Let us be given
\begin{prooftree}
\AXC{$\begin{aligned}
\Gamma &\vdash A:*\\
\Gamma &\vdash A_1:*\\
\Gamma &\vdash A_* : A \ee{} A_1
\end{aligned}$}
\AXC{$\begin{aligned}
\Gamma, x:A &\vdash B:*\\
\Gamma, x_1:A_1 &\vdash B_1:*\\
\Gamma, x:A, x_1:A_1, x_*: x \sim_{A_*} x_1 &\vdash B_* : B \ee{} B_1
\end{aligned}$}
\BIC{$\Sigma^*\, [x,x_1,x_*] : A_*.\, B_* : \Sigma x{:}A.B \ee{} \Sigma x_1{:}A_1.B_1$}
\end{prooftree}
We are to check that $(\Sigma^*\, [x,x_1,x_*] : A_*.\, B_*)^*$ has
type
\begin{align*}
&(\Sigma^*\, [x,x_1,x_*] : A_*.\, B_*) \sim_{(\Sigma x{:}A.B \ee{}
  \Sigma x_1{:}A_1.B_1)^*}
(\Sigma^*\, [x,x_1,x_*] : A_*.\, B_*)'\\
&=(\Sigma^*\, [x,x_1,x_*] : A_*.\, B_*) \sim_{\seq (\Sigma x{:}A.B)^* 
  (\Sigma x_1{:}A_1.B_1)^*}
(\Sigma^*\, [x',x'_1,x'_*] : A'_*.\, B'_*)\\
&= \prod \tripar{p : \Sigma x{:}A.B} {p' : \Sigma
  x'{:}A'.B'} {p^* : p \sim_{(\Sigma x{:}A.B)^*} p'}
\prod \tripar {p_1 : \Sigma x_1{:}A_1.B_1} {p_1' : \Sigma x_1'{:}A_1'.B_1'}
{p_1^* : p_1 \sim_{(\Sigma x_1{:}A_1.B_1)^*} p_1'}.\\
&\qquad\qquad (p \sim_{\Sigma^*\, [x,x_1,x_*] : A_*.\, B_*} p_1) \eeq
    (p' \sim_{\Sigma^*\, [x',x'_1,x'_*] : A'_*.\, B'_*} p'_1)
\end{align*}

Laying down this type on top of
\begin{align*}
  &(\Sigma^* [x,x_1,x_*] : A_*.\ B_*)^*\\
&\eq \blam \tripar{p : \Sigma x{:}A.B}{p':\Sigma x'{:}A'.B'}
{p^* : p \sim_{\Sigma^* A^* B^*} p'}
\blam \tripar {p_1 : \Sigma x_1{:}A_1.B_1} {p_1' : \Sigma x_1'{:}A_1'.B_1'} {p_1^* : p_1 \sim_{\Sigma^* A_1^* B_1^*} p_1'}.\\
&\phan
\sum\hista\trip{a_* : \pi_1 p \sim_{A_*} \pi_1 p_1}
{a'_* : \pi_1 p' \sim_{A'_*}  \pi_1 p'_1}
{a^*_* : a_* \sim_{(A_*^\sim\, \pi\!{}_1\! p\; \pi\!{}_1\! p\!{}_1)^*} a'_*} :
A^*_* \tripar{\pi_1 p}{\pi_1 p'}{\pi_1 p^*}
\tripar{\pi_1 p_1} {\pi_1 p_1'} {\pi_1 p_1^*}.\\
&\phanq B^*_* \trip{\pi_1 p/x}{\pi_1 p'/x'}{\pi_1 p^*/x^*}
\trip{\pi_1 p_1/x_1}{\pi_1 p_1'/x_1'}{\pi_1 p_1^*/x_1^*}
\trip{a_*/x_*}{a_*'/x_*'}{a_*^*/x_*^*}
\tripar{\pi_2 p}{\pi_2 p'}{\pi_2 p^*}
\tripar{\pi_2 p_1}{\pi_2 p_1'}{\pi_2 p_1^*}
\end{align*}
one can discern that the $\Pi$- and $\lambda$-binders have
similar domains.

The terms will forever be united in a valid typing judgment if
\begin{align*}
&\sum\hista\trip{a_* : \pi_1 p \sim_{A_*} \pi_1 p_1}
{a'_* : \pi_1 p' \sim_{A'_*}  \pi_1 p'_1}
{a^*_* : a_* \sim_{(A_*^\sim \, \pi\!{}_1\! p\; \pi\!{}_1\! p\!{}_1)^*} a'_*} :
A^*_* \tripar{\pi_1 p}{\pi_1 p'}{\pi_1 p^*}
\tripar{\pi_1 p_1} {\pi_1 p_1'} {\pi_1 p_1^*}.\\
&\phanq B^*_* \trip{\pi_1 p/x}{\pi_1 p'/x'}{\pi_1 p^*/x^*}
\trip{\pi_1 p_1/x_1}{\pi_1 p_1'/x_1'}{\pi_1 p_1^*/x_1^*}
\trip{a_*/x_*}{a_*'/x_*'}{a_*^*/x_*^*}
\tripar{\pi_2 p}{\pi_2 p'}{\pi_2 p^*}
\tripar{\pi_2 p_1}{\pi_2 p_1'}{\pi_2 p_1^*}
\end{align*}
has type
\begin{align*}
&(p \sim_{\Sigma^*\, [x,x_1,x_*] : A_*.\, B_*} p_1) \eeq
    (p' \sim_{\Sigma^*\, [x',x'_1,x'_*] : A'_*.\, B'_*} p'_1)\\
&\eq \left(\sum \hista \trip{x}{x_1}{x_*} : A_*.B_*\right)^\sim \!\!p\; p_1
\ \ \eeq \  \
\left(\sum \hista \trip{x'}{x'_1}{x'_*} : A'_*. B'_*\right)^\sim \!p'\;
p_1'\\
&\eq \hspace{-0.6cm}\sum_{a_* : \pi_1 p \sim_{A_*} \pi_1 p_1} \hspace{-0.3cm} B_*\trip{\pi_1
    p/x}{\pi_1 p_1/x_1}{a_*/x_*}^\sim \!\pi_2 p\;\pi_2 p_1
\quad \eeq \hspace{-0.3cm}
\sum_{a'_* : \pi_1 p' \sim_{A_*'} \pi_1 p'_1} \hspace{-0.3cm} B'_*\trip{\pi_1
  p'/x'}{\pi_1 p'_1/x'_1}{a'_*/x'_*}^\sim \!\pi_2 p'\;\pi_2 p'_1
\end{align*}

This is indeed the case, for by induction it so happens that
\begin{align*}
  A^*_* \tripar{\pi_1 p}{\pi_1 p'}{\pi_1 p^*}
\tripar{\pi_1 p_1} {\pi_1 p_1'} {\pi_1 p_1^*}
&\of (A_*^\sim \ \pi_1 p \ \pi_1 p_1) \eeq ({A'_*}^\sim\ \pi_1 p'
\ \pi_1 p_1')
\end{align*}
and, for $a^*_* : A^*_* \tripar{\pi_1 p}{\pi_1 p'}{\pi_1 p^*}
\tripar{\pi_1 p_1} {\pi_1 p_1'} {\pi_1 p_1^*}^\bsim a_*\ a_*'$,
\begin{align*}
& B^*_* \trip{\pi_1 p/x}{\pi_1 p'/x'}{\pi_1 p^*/x^*}
\trip{\pi_1 p_1/x_1}{\pi_1 p_1'/x_1'}{\pi_1 p_1^*/x_1^*}
\trip{a_*/x_*}{a_*'/x_*'}{a_*^*/x_*^*}
\tripar{\pi_2 p}{\pi_2 p'}{\pi_2 p^*}
\tripar{\pi_2 p_1}{\pi_2 p_1'}{\pi_2 p_1^*}\\
&\of 
\sim B_*\trip{\pi_1
    p/x}{\pi_1 p_1/x_1}{a_*/x_*} \;\pi_2 p\;\pi_2 p_1
\quad \eeq \quad
\sim  B'_*\trip{\pi_1
  p'/x'}{\pi_1 p'_1/x'_1}{a'_*/x'_*} \;\pi_2 p'\;\pi_2 p'_1
\end{align*}
This completes the case of $\Sigma$-congruence.
\item[$\eeq$-congruence] The last case left standing is the $\seq$-constructor:
\begin{prooftree}
\AXC{$\begin{aligned}
&\phabra{\Gamma \vdash A:\sta}\\
&\phabra{\Gamma \vdash A_1:\sta}\\
&\phabra\Gamma \vdash A_* : A \ee{} A_1
\end{aligned}$}
\AXC{$\begin{aligned}
&\phabra{\Gamma \vdash B:\sta}\\
&\phabra{\Gamma \vdash B_1:\sta}\\
&\phabra\Gamma \vdash B_* : B \ee{} B_1
\end{aligned}$}
\BIC{$\ee{}^* A_* B_* : (A \ee{} B) \ee{} (A_1 \ee{} B_1)$}
\end{prooftree}
To get it down, we just need to force
\begin{align*}
 (\seq A_* B_*)^* &\eq \blam \tripar{e : A \eeq B}
{e':A' \eeq B'} {e^* : e \sim_{\seq A^* B^*} e'}
\quad\blam\tripar{e_1 : A_1 \eeq B_1}{e_1' : A_1' \eeq B_1'}
{e_1^* : e_1 \sim_{\seq A_1^* B_1^*} e_1'}.\\
&\phanq
\prod\hista \trip{a}{a'}{a^*} : A^*\;
\prod\hista \trip{a_1}{a_1'}{a_1^*} : A_1^*\;
\prod\hista \trip{a_*}{a_*'}{a_*^*} : A_*^* \tripar{a}{a'}{a^*} \tripar{a_1}{a_1'}{a_1^*}\\
&\phanq
\prod\hista \trip{b}{b'}{b^*} : B^*\;
\prod\hista \trip{b_1}{b_1'}{b_1^*} : B_1^*\;
\prod\hista \trip{b_*}{b_*'}{b_*^*} : B_*^* \tripar{b}{b'}{b^*} \tripar{b_1}{b_1'}{b_1^*}.\\
&\phanq\qquad \bseq \left(e^* \tripar{a}{a'}{a^*}
  \tripar{b}{b'}{b^*}\right)
\left(e_1^* \tripar{a_1}{a_1'}{a_1^*} \tripar{b_1}{b_1'}{b_1^*}\right)\\
\end{align*}
into
\renewcommand{\phan}{\hspace{1.17cm}}
\begin{align*}
&\phan \seq A_* B_* \sim_{((A \eeq B) \eeq (A_1 \eeq B_1))^*} \seq A_*' B_*'\\
&\eq \seq A_* B_* \sim_{\seq (A \eeq B)^* \eeq (A_1 \eeq B_1)^*} \seq A_*' B_*'\\
&\eq \prod \tripar {e : A \eeq B}{e':A' \eeq B'}{e^* : e \sim_{(A
  \eeq B)^*} e'} \prod \tripar {e_1 : A_1 \eeq B_1}{e_1':A_1' \eeq B_1'}{e_1^*: e_1 \sim_{(A_1
  \eeq B_1)^*} e_1'}.\\
&\phanq\quad (e \sim_{\seq A_* B_*} e_1) \eeq (e' \sim_{\seq A_*' B_*'} e_1')
\end{align*}
\newcommand{\beq}{\scalebox{1.7}{${\eeq}$}}
The lambdas go into the pies quite easily, so we focus on
\begin{align}
&(e \sim_{\seq A_* B_*} e_1) \eeq (e' \sim_{\seq A_*' B_*'} e_1')
\eq \label{yeaaay}\\
&\phanq \qquad \left(\begin{aligned}
&\Pi a{:}A \Pi a_1{:}A_1 \Pi a_* : a \sim_{A_*} a_1\\
&\Pi b{:}B \Pi b_1{:}B_1 \Pi b_* : b \sim_{B_*} b_1. \quad
 a \sim_e b \ \eeq\ a_1 \sim_{e_1} b_1
\end{aligned}\right) \notag\\
&\phanq \quad\beq \left(\begin{aligned}
&\Pi a'{:}A' \Pi a'_1{:}A'_1 \Pi a'_* : a' \sim_{A'_*} a'_1\\
&\Pi b'{:}B' \Pi b'_1{:}B'_1 \Pi b'_* : b' \sim_{B'_*} b'_1. \quad
 a' {\sim_{e'}} b' \ \eeq\ a'_1 {\sim_{e'_1}} b'_1
\end{aligned}\right) \notag
\end{align}
To inhabit this equivalence type, one needs to construct a sequence
of equivalences which pairwise relate the domains of quantification
in the sequence of $\Pi$-types on each side of the $\eeq$-sign.

Close inspection will reveal that the six $\Pi^*$-constructors
appearing in the unfolding of
$(\seq A_* B_*)^*$ do provide such a sequence.
For example, by induction hypothesis, we have
\[A_*^* : \prod \tripar{x : A}{x': A'}{x^*: x \sim_{A^*} x'}
\prod \tripar{x_1:A_1}{x_1':A'_1}{x_1^* : x_1 \sim_{A_1^*} x_1'}.\ 
x \sim_{A_*} x_1 \; \eeq \; x' \sim_{A'_*} x'_1\]
whence we get the equivalence
\[ A_*^* \tripar{a}{a'}{a^*} \tripar{a_1}{a_1'}{a_1^*}  \of
a \sim_{A_*} a_1 \ \ \eeq\ \  a' \sim_{A'_*} a'_1\]
relating the domains of the third $\Pi$s in the sequence.

(Similarly, $\ B_*^* b b' b^* b_1 b_1' b_1^* \; : \;
b \sim_{B_*} b_1 \;\eeq\; b' \sim_{B'_*} b'_1 $.)

All that remains is to check that
\begin{align*}
&\bseq \left(e^* \tripar{a}{a'}{a^*}
  \tripar{b}{b'}{b^*}\right)
\left(e_1^* \tripar{a_1}{a_1'}{a_1^*}
  \tripar{b_1}{b_1'}{b_1^*}\right)\\
&\of 
(a \sim_e b \ \eeq\ a_1 \sim_{e_1} b_1) \ \eeq \
(a' \sim_{e'} b' \ \eeq\ a'_1 \sim_{e'_1} b'_1)
\end{align*}

Indeed, this is attainable from
\begin{align*}
  e^* a a' a^* b b' b^* &\of (a \sim_e b) \eeq (a' \sim_{e'} b')\\
  e_1^* a_1 a_1' a_1^* b_1 b_1' b_1^* &\of
(a_1 \sim_{e_1} b_1) \eeq (a_1' \sim_{e_1'} b_1')
\end{align*}
by feeding these terms into an application of the $\seq$-rule.
\end{description}
This completes the proof of the theorem.
\end{proof}

\section{Extensional equality of ground types}
\label{s:exte}

We observe some consequences of the theorem.

\begin{enumerate}
\item Consider the type judgement
  \[ \vdash \sta : \sta \]
It is derivable by an axiom; by applying the theorem, we get
\[ \vdash \sta^* : \sta \sim_{\sta^*} \sta \]
By the reduction rule for $\eeq$,
\[ \sta \sim_{\sta^*} \sta \rrule \sta \eeq \sta \]
By conversion rule, the theorem is thus saying that
\[ \vdash \sta^* : \sta \eeq \sta \]
which is indeed the case (axiom).
\item 
Consider a type judgement
\[ \vdash A : \sta \]
Applying the theorem gives
\[ \vdash A^* : A \sim_{\sta^*} A' \]
By conversion rule, we have $A^* : A \eeq A'$.

But $A$ is closed term.  So every variable of $A$ is bound.

$A'$ is obtained from $A$ by apostrophizing every variable.

So $A'$ is alpha equivalent to $A$.

And the type of $A^*$ is alpha equivalent to $A \eeq A$.

\begin{definition}
  Let $\vdash A : \sta$ be a closed type.  We define \emph{extensional
    equality on $A$} to be \[\sim_{A^*} \of A \to A \to \sta\]
\[    \fbox{ $a \ee A a' \df a \sim_{A^*} a'$ }\]
\end{definition}
$A^*$ may be called 
\emph{trivial type equality}, \emph{identity equivalence} (on $A$),
\emph{reflexivity of $A$}.

\item Consider a type judgement
\[ \vdash a :A\]
Applying the theorem gives
\[ \vdash a^* : a \sim_{A^*} a' \]
Using the previous definition, we write this as
\[ \vdash a^* : a \ee A a' \]
But $a$ is closed term.  So every variable of $a$ is bound.

$a'$ is obtained from $a$ by apostrophizing every variable.
So $a'$ is $\alpha$-equal to $a$.

And the type of $a^*$ is $\alpha$-equal to $a \ee A a$.

\begin{definition}
Let $\vdash a : A$ be a closed term.  We define the \emph{reflexivity
  of $a$} to be \[a^* \of a \ee A a\]
\[ \fbox{$ \refl{a} \df a^* $}\]
\end{definition}
\item
A particular case of the above is the judgment $ \vdash A : \sta $

As before, we derive $\refl{A} = A^\sta : A \ee \sta A$.  Then
\[ a \ee A a' \eq a \sim_{\refl{A}} a' \]

For closed terms, the following rule is thus derived:
\[
\fbox{$\AXC{$ \vdash a : A $}
\UIC{$ \vdash \refl{a} : a \sim_{\refl{A}} a$}
\DisplayProof$}
\]
 
In particular, we derive $\refl{\sta} = \sta^\sta : \sta \ee \sta \sta$.  Then
\[ A \ee \sta B \eq A \sim_{\refl{\sta}} B \eq A \eeq B \]
\[ \fbox{$A \ee \sta B \eq A \eeq B $}\]

\item
Consider a type judgment
\[ \vdash a : A \]
Using the theorem, we derive
\[ \vdash \refl{a} : a \ee A a \]
Applying the theorem again gives
\begin{align*}
&\vdash \refl{\refl{a}} \quad\; \of \refl{a} \ee {a \ee A a}
\refl{a}\\
&\vdash \refl{\refl{\refl{a}}} \of \refl{\refl{a}} \ee {\refl{a} \ee {a \ee A a}
\refl{a}}
\refl{\refl{a}}\\
&\qquad \vdots
\end{align*}

\item Suppose we have derivations
  \begin{align*}
    x : A &\vdash b(x) : B(x)\\
    &\vdash a^* : a \ee A a'
  \end{align*}

Applying the theorem, we get
\begin{align*}
 x:A, x':A, x^* : x \sim_{\refl{A}} x' &\vdash B^*(x,x',x^*) : B(x)
 \eeq B(x')\\
 x:A, x':A, x^* : x \sim_{\refl{A}} x' &\vdash b^*(x,x',x^*) : b(x) \sim_{B^*(x,x',x^*)}
b(x')
\end{align*}
In particular, we obtain
\begin{align*}
 &\vdash B^*(a,a',a^*) : B(a) \eeq B(a')\\
 &\vdash b^*(a,a',a^*) : b(a) \sim_{B^*(a,a',a^*)} b(a')
\end{align*}

\item  
If we furthermore have
\[ x:A, y:B(x) \vdash c(x,y) : C(x,y) \]
then we also obtain
\begin{align*}
 &\vdash C^*(a,a',a^*,b(a),b(a'),b^*(a,a',a^*)) : C(a,b(a)) \eeq C(a',b(a'))\\
 &\vdash c^*(\vec a, \ora{b(a)}) : c(a,b(a)) \sim_{C^*(\vec a,
   \ora{b(a)})} c(a',b(a'))
\end{align*}
\end{enumerate}

\begin{definition}
Let $\Gamma = (x_1 : A_1, \dots, x_n : A_n(x_1,\dots,x_{n-1}))$ be a
context.  A \emph{path in $\Gamma$} is a sequence of terms
\[ (\vec a, \vec a', \vec a^*) \eq (a_1,a_1',a_1^*,\dots,a_n,a_n',a_n^*) \]
such that, for each $i \in \setof{0,\dots,n-1}$, the following holds:
\begin{align*}
  &\vdash a_i : A_i(a_1,\dots,a_{i-1})\\
  &\vdash a_i' : A_i(a_1',\dots,a_{i-1}')\\
  &\vdash a_i^* : a_i
  \sim_{A_i^*(a_1,a_1',a_1^*,\dots,a_{i-1},a_{i-1}',a_{i-1}^*)} a_i'
\end{align*}
When $(\vec a, \vec a', \vec a^*)$ is a path in $\Gamma$, we write
\[ \vec a^* \of \vec a \ee \Gamma \vec a' \]
\end{definition}

{\sc Corollary.}  The theorem of Section \ref{ext-thm}
has the following consequences.
\begin{itemize}
\item Every ground type possesses the structure of a
  globular set with degeneracies.
\item If $\Gamma \vdash B(x_1,\dots,x_n) : \sta$, then every path
  $(\vec a, \vec a', \vec a^*)$ in $\Gamma$ induces a type equality
\[ \vdash B(\vec a^*) : B(\vec a) \eeq B(\vec a') \]
\item If $\Gamma \vdash b(\vec x) : B(\vec x)$, then every path
$(\vec a, \vec a', \vec a^*)$ in $\Gamma$ induces an equality over
$B^*(\vec a, \vec a', \vec a^*)$:
\[ \vdash b(\vec a^*) : b(\vec a) \sim_{B(\vec a^*)} b(\vec a') \]
\end{itemize}
\section{Stratification and semantics}
\label{strat}

\newcommand{\leeqn}{\leeq_n}
\subsection{Stratification of $\leeq$}
\begin{definition}
The system $\leeqn$ is obtained from $\leeq$ by executing the following recipe:
\begin{enumerate}
\item The symbol $\sta$ is replaced by an infinite collection of
  constants
\[\setof{\sta_n \mid n \in \omega}\]
\item The typing rule $\ol{ \vdash \sta : \sta}$ is replaced by the
  rule scheme (one rule for each number $n$):
\[\AXC{$$} \UIC{$\vdash \sta_n : \sta_{n+1}$}
  \DisplayProof\]
\item A new rule is introduced:
  \begin{prooftree}
    \AXC{$\Gamma \vdash A : \sta_n$}
    \UIC{$\Gamma \vdash A : \sta_{n+1}$}
  \end{prooftree}
\item In all other rules,
the $\sta$ symbol is replaced by $\sta_n$.
\end{enumerate}
\end{definition}
 
{\sc Remark.}\ \ When we describe the intended model, it will turn out that the above
definition is not quite correct: the elimination rule for $\eeq$ does
not have the right universe indexing.  We shall address this issue when
it arises in the course of our construction.

\subsection{The strict model}

First, we describe a particularly simple model in which $A \eeq B$ is interpreted by
strict, set-theoretic equality.  This model even validates the rule
derivable in the \cite{itt} system with propositional reflection:
\[ \AXC{$\Gamma \vdash a : A$}
\AXC{$\Gamma \vdash e : A \eeq B$}
\BIC{$\Gamma \vdash a : B$}
\DisplayProof \]

Let $\kappa_0 \subset \kappa_1 \subset \cdots$ be a sequence
of strongly inaccessible cardinals.

\begin{enumerate}
\item Each universe $\sta_n$ is interpreted by $\bset_n =
  V_{\kappa_n}$, the cumulative hierarchy up to stage~$\kappa_n$:
\[ \semof{\sta_n} \df \bset_n \]
\item The $\Pi$- and $\Sigma$-types are interpreted,
  respectively, by cartesian product and disjoint union of families of sets:
\begin{align*}
    \semof{\Pi x{:}A.B(x)}
    &\df \prod_{a \in \semof{A}} \semof{B}_{x:=a}
 &&= \quad \setof{ f : \semof{A}
      \to \bigcup_{a \in \semof{A}} \semof{B}_{x:=a} \mid 
      \forall a.\ f a \in \semof{B}_{x:=a}}\\
    \semof{\Sigma x{:}A.B(x)} &\df \bigsqcup_{a \in \semof{A}}
    \semof{B}_{x:=a}
&&= \quad \setof{(a,b) \mid a \in \semof{A}, b \in
      \semof{B}_{x:=a}}
  \end{align*}
\item The $\eeq$-type is interpreted by equality:
\[ \semof{A \eeq B} \df \begin{cases}
\setof{\emptyset} &\semof{A} = \semof{B}\\
\emptyset &\semof{A} \neq \semof{B}
\end{cases} \]
\end{enumerate}

Since, for $\kappa$ strongly inaccessible,
$V_\kappa$ is closed under cartesian products and disjoint union,
the above definition manifestly
validates the four formation rules of $\lstarn$, as well as the
subsumption rule.

The interpretation of term formers related to the $\Pi$- and
$\Sigma$-types is completely standard:
\begin{align*}
    \semof{\lambda x{:}A.t}\rho &\eq (a \mapsto
    \semof{t}_{\rho,x:=a})  \qquad\qquad {\in \prod{}_{a{\in}\semof{A}_\rho}\semof{B}_{\rho,a}}\\
    \semof{f a}_\rho &\eq \semof{f}_\rho(\semof{a}_\rho)\\
    \semof{(a,b)}_\rho &\eq (\semof{a}_\rho,\semof{b}_\rho)
    \hspace{1.87cm}{\in \bigsqcup{}_{a{\in}\semof{A}_\rho} \semof{B}_{\rho,a}}\\
    \semof{\pi_i t}_\rho &\eq p_i \quad \text{ where }(p_1,p_2)=\semof{t}_\rho
\end{align*}

The interpreation of $\Pi^*, \Sigma^*, \seq, \sta^\sta$ is
self-evident.  When $e : A \eeq B$, put
\[  \semof{ a \sim_e b} \df \begin{cases}
\setof{\emptyset} & \semof{a} = \semof{b}\\
\emptyset &\text{otherwise}
\end{cases}\]

The interpretation of contexts $\Gamma$ is the set of all tuples
$(a_1,\dots,a_n)$ such that
\[ a_{i+1} \in \semof{A_{i+1}}_{a_1,\dots,a_i}
\qquad (0 \le i < n) \]

It is straightforward to verify that the interpretation preserves
substitution, conversion, and typing rules.  So we have

\begin{theorem} (Soundness)
Let $\semof{\Gamma}, \semof{A}_{\rho \in \semof{\Gamma}},
\semof{a}_{\rho \in \semof{\Gamma}}$ be as defined above.  Then
\[ \Gamma \vdash M : A \qquad \then \qquad
\semof{M} : \prod_{\vec a \in \semof{\Gamma}} \semof{A}_{\vec a} \]
\end{theorem}

The above model is \emph{proof-irrelevant}, since proofs of equality
have no computational content.  It is in keeping with our goal of
generality, that extensional equality should admit such an
interpretation.

However, the $\eeq$-type also contains all the necessary machinery
for transporting computational information over proofs of equality.
We shall now describe a model which makes use of this feature.

Due to lack of space, we do not go into details, but give a general outline.
\subsection{The proof-relevant model}

\subsubsection{Isomorphism as equality}

In this model, type equality is interpreted as isomorphism of sets:
\[ \semof{A \eeq B}  \df \semof{A} \eeq \semof{B} \]

Let $\setof{B_x \mid x \in A}$, $\setof{B'_y \mid y \in A'}$ be
families of sets.
Given an isomorphism $i : A \stackrel{\eeq}{\to} A'$,
and a family $\setof{j_{a,a'} : B_a
  \stackrel{\eeq}{\to} B_{a'} \mid i(a) = a'}$, we obtain isomorphisms

\begin{align*} \Pi^*_{i(x)=y} (j_{x,y}) &\of \prod_{x \in A} B_x
  \ \ \stackrel{\eeq}{\rrule} \ \ \prod_{y \in A'} B'_y\\
\sqcup^*_{i(x)=y}(j_{x,y}) &\of \bigsqcup_{x \in A} B_x
\ \ \stackrel{\eeq}{\rrule}\ \ \bigsqcup_{y \in A'} B'_y
\end{align*}

Given isomorphisms $i : A \iso A'$, $j : B \iso B'$, we get an
isomorphism (conjugation):
\begin{align*}
 \seq(i,j) \of (A \eeq B) &\rrule  (A' \eeq B') \\
 \seq(i,j) \of \hspace{1.06cm}\xi &\quad\longmapsto\quad j \circ \xi \circ i^{-1}
\end{align*}

Every isomorphism $i : A \eeq B$ induces a binary relation
$\tilde i \subseteq A \times B$:
\begin{align*}
a \tilde i b &\quad\iff\quad i(a) = b
\end{align*}

Every set $A$ has the identity isomorphism:
\[ \mathsf{I}_A : A \eeq A \]

In particular, there exist canonical isomorphisms
\[ \mathsf{I}_{\bset_n} \of V_{\kappa_n} \eeq V_{\kappa_n} \]

This fixes the interpretation of everything related to type equality.
Keeping the interpretation of other types the same, we now try to
validate the reduction rules.

\subsubsection{A bug?}

Thankfully, all of the reduction rules are perfectly valid in our model.

Except one.
\begin{equation}
 A \sim_{\sta^*} B \rrule A \eeq B \label{srule}
\end{equation}

The right side is interpreted by the set of isomorphisms $\semof{A}
\eeq \semof{B}$.

The left side is interpreted by the relation $\tilde{\mathsf{I}}_{\bset_n}
\subseteq \bset_n \times \bset_n$:
\[ \semof{A \sim_{\sta^*} B}
\eq \begin{cases}
\setof{\emptyset} & \semof{A} = \semof{B}\\
\emptyset &\semof{A} \neq \semof{B}
\end{cases} \]

The rule \eqref{srule}
is thus saying that there is at most one isomorphism between
any pair of sets, a claim which many members of $\bset_n$ will find offensive.

How are we to reconcile this reduction rule with our model?

\subsubsection{The truth-table universe}

In the first instance, we notice that the
offending claim \emph{does} make sense
for sets that are either empty or singletons, i.e., 
\emph{propositions}.

Indeed, the universe $\setof{0,1}$ of classical propositions is closed under
isomorphism, in the sense that there is indeed at most one isomorphism
between any two propositions --- its existence being equivalent to the
existence of a pair of maps between them.
 \comment{ (The relation $\tilde e$
induced by such an isomorphism $e$ 
relates all pairs of elements, and is equivalent to the
existence of functions from one set to the other and back.
(In particular, the identity relation on an empty or singleton set relates
all elements of that set.)}

Furthermore, all other type constructors can be given the standard
``truth-table semantics'' in this universe,
validating their introduction and elimination rules.

This motivates us to let propositions actualize the
interpretation of $\sta_0$, the lowest universe in $\leeqn$.
Constructions carried out in this universe fall in the scope
of the \emph{propositions as types} embedding (\cite{Howard80}).

\subsubsection{Homotopy hierarchy}

We are thus led to reconsider our interpretation
\[\semof{\sta_n} = \bset_n, \qquad
\semof{\sta_n^*} = \bset_n^* = \mathsf{I}_{\bset_n} : \bset_n \iso \bset_n \]

In order for \eqref{srule} to remain valid while preserving
``type equality is isomorphism'' idea, we \emph{must} observe
\[ \semof{A} \widetilde{\bset^*} \semof{B}
\eq \semof{A} \eeq \semof{B} \]

This means that the equality on $\bset$ --- which we defined as the
``relation'' induced by identity equality of $\bset$ with itself ---
must actually be a \emph{set family} (giving, for any two sets,
the set of isomorphisms between then), rather than a simple relation.

In contrast, for $A, B$ elements of $\bset$,
the relation $\tilde e \subseteq A \times B$ induced by 
an isomorphism  $e : A \iso B$, is always two-valued: it's a proposition.
In particular, the equality relation induced by the identity isomorphism on
$A$ is a proposition.

For propositions, an ``isomorphism'' is just a pair of maps, and the
relation associated to this pair is the total (``1-valued'') relation between the two propositions.

Going higher, we find that, for groupoids $G_1, G_2$,
the collection of groupoid equivalences
between $G_1$ and $G_2$ forms again a groupoid\footnote{
Given $E, E' : G_1 \stackrel{\eeq}{\RA} G_2$, 
the isomorphisms between $E$, $E'$ may be given equivalently either as
\[  \prod_{A \in G_1} \prod_{B \in G_2}
G_2(E(A),B) \eeq G_2(E'(A),B) 
\qquad \text{or} \qquad \prod_{A \in G_1} G_2(E(A),E'(A))\]
These collections, being products of sets, are again sets.},
 and every groupoid equivalence  $E : G_1 \RA G_2$ induces a
$\bset$-valued predicate on $G_1 \times G_2$:
\[ \tilde{E} \df (A \in \objof{G_1}) \mapsto (B \in \objof{G_2})
\mapsto \homsym_{G_2}(E(A),B) \]

We observe the following pattern:\\

{\centering \fbox{\begin{minipage}{10cm}
{The relation ${\sim}e : A \to B \to \sta$ induced by a type equality
$e : A \eeq B$ between types in the universe $\sta_n$,
is valued in the universe $\sta_{n-1}$.}
\end{minipage} }\\ \par} \vspace{0.3cm}

This pattern leads us to revoke the interpretation of universes in the
cumulative hierarchy of set theory in favor of the (still cumulative)
hierarchy of homotopy $n$-types:
\begin{align*}
\vspace{0.2cm}\semof{ \sta_0 \le \sta_1 \le \sta_2 \le \cdots \le \sta_n \le
  \cdots} &\df 
\mathsf{Prop} \subseteq \mathsf{Set} \subseteq \mathsf{Grpd}
\subseteq \cdots \subseteq {(n{-}2)}{\text{-}}\mathsf{Grpd} \subseteq
\cdots \\
&\,\eq\, \mathsf{Prop} \in \mathsf{Set} \in \mathsf{Grpd} \in
\:\cdots \:\in {(n{-}2)}{\text{-}}\mathsf{Grpd} \in \cdots
\vspace{0.4cm}\end{align*}

\subsubsection{Fixing the bug}
 
The pattern announced above forces us to reconsider the
$\eeq$-elimination rule in the
stratified system.  It shall now be read as follows.

\begin{prooftree}
  \AXC{$A : \sta_n$}
  \AXC{$e : A \eeq B$}
  \AXC{$B : \sta_n$}
  \TIC{${\sim}e : A \to B \to \sta_{n-1}$}
\end{prooftree}

Postponing for the moment the question of what we are to make of the
conclusion in the case when $n=0$, we point out that this change
resolves the problem in $\eqref{srule}$, allowing us to complete
the model.  Thus, for $A,B : \sta_{n}$, we have
\begin{align*}
\semof{A \eeq B} &=
\text{ {$n$}-equivalence between $n$-types,}\\
\semof{A \sim_{\sta_{n}^*} B} &=
\text{ $n$-relation induced by the identity equivalence of the
$(n{+}1)$-type $\sta_{n}$.}
\end{align*}
These collections may be naturally identified.  In particular:\\
- Two propositions are equivalent if their truth-table semantics yield
isomorphic sets;\\
- Two sets $A, B$ are isomorphic if the identity groupoid equivalence
$\mathsf{I}_\bset : \bset \stackrel{\eeq}{\RA} \bset$
relates them as objects: $(A \eeq B) = \bset(\mathsf{I}_\bset(A),B)$.\\
- etc.

We remark that the formation rule of the equality type needs
no amendments:
\[
\AXC{$A : \sta_n$}
\AXC{$B : \sta_n$}
\BIC{$A \eeq B : \sta_n$}
\DisplayProof
\]
This rule already gives the intended meaning:\\
- Propositions are closed under logical equivalence;\\
- Sets are closed under isomorphism;\\
- Groupoids are closed under equivalence of groupoids;\\
- etc.

Although we have discussed only the first few levels of the homotopy
hierarchy, it is clear that the given pattern has a clear inductive
structure, and extends to all finite $n$-types.

We may also consider adding a ``limit universe''
\[ \sta_0 \le \sta_1 \le \sta_2 \le \sta_3 \le \cdots \qquad
\le \sta_\omega \]
for which the ${\sim}(\cdot)$-operator would stay valued in $\sta_\omega$.  The
natural interpretation of $\sta_\omega$ would be by a model of weak
$\omega$-groupoids.

But in order for such a universe to be of any interest, our language
should already provide the computational interpretation of the
higher groupoid laws (the Kan filling conditions).
This of course is a major topic for future work.

\newcommand{\unitt}{\boldsymbol{1}}
\newcommand{\unitc}{t\!t}
\newcommand{\unitelim}{\mathsf{Const}}
To complete our model, we discuss the rule of $\eeq$-elimination for
the case when $n = 0$.

The symbol $\sta_{-1}$ is treated as notation for $\unitt$, the unit
type.

We add new formation, introduction, elimination, and computation
rules for this type.

\[  \AXC{}
  \UIC{$\Gamma \vdash \unitt : \sta_0$}
  \DisplayProof
  \qquad \quad
  \AXC{}
  \UIC{$\Gamma \vdash \unitc : \unitt$}
  \DisplayProof\]
\vspace{0.1 cm}
\[
  \AXC{$\Gamma, x : \unitt \vdash B : \sta$}
  \AXC{$\Gamma \vdash b : B[\unitc/x]$}
  \BIC{$\Gamma \vdash \unitelim_B(b) : \Pi x{:} \unitt. B $}
  \DisplayProof \]
\vspace{0.3 cm}
\[
  \unitelim_B(b) \unitc \rrule b \]

We add the rules for interaction between $\unitt$ and
${\eeq}$.
\begin{prooftree}
  \AXC{}
  \UIC{$\Gamma \vdash \unitt^* : \unitt \eeq \unitt$}
\end{prooftree}
\[ x \sim_{\unitt^*} y \rrule \unitt \]
Finally, we extend the $(\cdot)^*$-operation to these new terms:\footnote{
We do not include the lifting of the $\unitelim$ eliminator; in future
versions of our system $\unitelim$ is expected to be derivable from the
\emph{transport operator}:
\begin{prooftree}
  \AXC{$ \Gamma, x \in \unitt \vdash B(x) : \sta$}
  \AXC{$ \Gamma \vdash u : \unitt$}
  \BIC{$\Gamma \vdash B^*(\unitc,u,\unitc) : B(\unitc) \eeq B(u)$}
  \UIC{$\Gamma \vdash B^*(\unitc,u,\unitc)^+ : B(\unitc) \to B(u)$}
  \AXC{$\Gamma \vdash b : B(\unitc)$}
  \BIC{$\Gamma \vdash B^*(\unitc,u,\unitc)^+ b : B(u)$}
\end{prooftree}
}
\begin{align*}
  (\unitt)^* &\eq \unitt^*\\
  (\unitc)^* &\eq \unitc \qquad (: \unitc \sim_{\unitt^*} \unitc) \\
  (\unitt^*)^* &\eq  \blam \tripar {x : \unitt}{x': \unitt} {x^* : x \ee
    \unitt x'}
\blam \tripar {y : \unitt} {y': \unitt} {y^* : y \ee \unitt y'}.\ \unitt^*
\end{align*}
 
This completes our model construction.

\section{Conclusion}

In this paper, we have enunciated Tait's suggestion for the
type-theoretic meaning of the notion of extensional equality.

We have shown how the external definition of extensional equality may
be reflected into the syntax.  Our construction yields an internal
definition of extensional equality for closed types.

We have not yet witnessed all of the desired properties of this
equality.  Future work includes generalization to open terms and
computational treatment of Kan filling conditions (to be defined via
transport maps over type equality).

In the model outlined in the last section, we see the syntactic
approach to extensional equality starting to come together
with the threads of ideas motivated by homotopy theory.

An original feature of this interpretation is that the logical
relation defined by induction on type structure is reflected in the
\emph{lower} universe than the types being related by it.

\bibliography{ett2.5}
\setlength{\voffset}{0cm}

\end{document}

%% file: amslambda.tex
\usepackage{amsmath}
\usepackage{stmaryrd}
\usepackage{MnSymbol}
\usepackage{bussproofs}
\EnableBpAbbreviations

\DeclareFontFamily{OT1}{pzc}{}
\DeclareFontShape{OT1}{pzc}{m}{it}{<-> s * [1.10] pzcmi7t}{}
\DeclareMathAlphabet{\mathpzc}{OT1}{pzc}{m}{it}

\newcommand{\termsof}[1]{\mathpzc{Terms}({#1})}

\newcommand{\leeq}{\lambda {\eeq}}

\newcommand{\lext}{\lambda\mathpzc{e}}

\newcommand{\then}{\;\Longrightarrow\;}
\newcommand{\comment}[1]{}
\newcommand{\RA}{\Rightarrow}

\newcommand{\lRa}{\longrightarrow}

\newcommand{\ora}{\overrightarrow}

\newcommand{\greyout}[1]{\textcolor{gray}{#1}}

\newcommand{\sta}{*}
\newcommand{\lstar}{{\lambda\!\!\:\sta}}

\newcommand{\qfun}{\mathbin{{\ooalign{$\medcircle$\cr\raise.29ex
\hbox{$\mkern1.9mu\scriptscriptstyle{\to}$}\cr}}}}

\newcommand{\ee}[1]{\simeq_{#1}}

\newcommand{\setof}[1]{\{{#1}\}}

\newcommand{\semof}[1]{\llbracket{#1}\rrbracket}

\newcommand{\refl}[1]{{\mathsf{r}({#1})}}

\newcommand{\ol}[1]{\overline{#1}}

\newcommand{\qFun}{\mathbin{{\ooalign{$\bigcircle$\cr\raise.037ex
\hbox{$\mkern3.1mu\scriptstyle{\Pi}$}\cr}}}}
\newcommand{\qSum}{\mathbin{{\ooalign{$\bigcircle$\cr\raise.036ex
\hbox{$\mkern3.5mu\scriptstyle{\Sigma}$}\cr}}}}

\newcommand{\qU}{\mathbin{{\ooalign{$\bigcircle$\cr\raise.037ex
\hbox{$\mkern5mu\scriptstyle{\mathtt{U}}$}\cr}}}}
\newcommand{\qT}{\mathbin{{\ooalign{$\bigcircle$\cr\raise.037ex
\hbox{$\mkern5mu\scriptstyle{\mathtt{T}}$}\cr}}}}
\newcommand{\qEq}{\mathbin{{\ooalign{$\bigcircle$\cr\raise.15ex
\hbox{$\mkern5mu\scriptstyle{\ee{}}$}\cr}}}}
\newcommand{\qRel}{\mathbin{{\ooalign{$\bigcircle$\cr\raise.2ex
\hbox{$\mkern5mu\scriptstyle{\sim}$}\cr}}}}

\newcommand{\idtype}{\mathpzc{Id}}

\newcommand{\thra}{\twoheadrightarrow}

\newcommand{\eq}{\quad=\quad}
\newcommand{\of}{\quad : \quad}
\newcommand{\de}{\quad := \quad}

\comment{

}

\newcommand{\phan}{\phantom{\eq}\;}
\newcommand{\phanq}{\phantom{\eq\qquad}}

\newcommand{\hista}{\!\raisebox{1mm}{${}^*$}}
\newcommand{\bsim}{\raisebox{-1mm}{\scalebox{1.5}{$\sim$}}}
\newcommand{\blam}{\raisebox{-2mm}{\scalebox{2.1}{$\lambda$}}}
\newcommand{\bseq}{\raisebox{-1mm}{\scalebox{1.8}{${\eeq^*}$}}}

\newcommand{\rrule}{\hspace{1em}\lRa\hspace{1em}}
\newcommand{\eeq}{\simeq}
\newcommand{\seq}{{\eeq}^*}

\newcommand{\ssim}{{\sim}}
\newcommand{\too}{\quad\longrightarrow\quad}

\newcommand{\optarg}[1]{{\textcolor{Periwinkle}{\setof{#1}}}}

\renewcommand{\refl}[1]{\mathsf{r}({#1})}

\newcommand{\trip}[3]{
{\scriptsize \left[
  \begin{array}{l}
    {#1}\\
    {#2}\\
    {#3}
  \end{array}
\right]}}

\newcommand{\tripar}[3]{
{\scriptsize \left(
  \begin{array}{l}
    {#1}\\
    {#2}\\
    {#3}
  \end{array}
\right)}}

\newcommand{\df}{\quad:=\quad}

\newcommand{\bset}{\mathsf{Set}}

\newcommand{\lstarn}{\lstar\!{}_n}

\newcommand{\ul}[1]{\underline{#1}}